\documentclass[letterpaper]{article} % DO NOT CHANGE THIS
\usepackage{aaai24}  % DO NOT CHANGE THIS
\usepackage{times}  % DO NOT CHANGE THIS
\usepackage{helvet}  % DO NOT CHANGE THIS
\usepackage{courier}  % DO NOT CHANGE THIS
\usepackage[hyphens]{url}  % DO NOT CHANGE THIS
\usepackage{graphicx} % DO NOT CHANGE THIS
\usepackage{natbib}  % DO NOT CHANGE THIS AND DO NOT ADD ANY OPTIONS TO IT
\usepackage{caption} % DO NOT CHANGE THIS AND DO NOT ADD ANY OPTIONS TO IT
\usepackage[noend]{algpseudocode}
\usepackage{algorithm}

\algnewcommand{\algorithmicwith}{\textbf{With}}
\algdef{SE}[FOR]{With}{EndWith}[1]{\algorithmicwith\ #1\ \algorithmicdo}{\unskip\ \algorithmicend\ \algorithmicwith}
\algtext*{EndWith}

\usepackage{amsmath}
\usepackage{dsfont}
\usepackage{subfigure}
\usepackage{amssymb}
\usepackage{booktabs}
\usepackage{bm}
\usepackage{multirow}

\usepackage{newfloat}
\usepackage{listings}
\DeclareCaptionStyle{ruled}{labelfont=normalfont,labelsep=colon,strut=off} % DO NOT CHANGE THIS
\floatstyle{ruled}
\newfloat{listing}{tb}{lst}{}
\floatname{listing}{Listing}
\title{Cautiously-Optimistic Knowledge Sharing\\for Cooperative Multi-Agent Reinforcement Learning}
\author {
    Yanwen Ba\textsuperscript{\rm 1},
    Xuan Liu\textsuperscript{\rm 1}\thanks{Corresponding Author.},
    Xinning Chen\textsuperscript{\rm 1},
    Hao Wang\textsuperscript{\rm 1},
    Yang Xu\textsuperscript{\rm 1},
    Kenli Li\textsuperscript{\rm 1},
    Shigeng Zhang\textsuperscript{\rm 2}
}
\affiliations {
    \textsuperscript{\rm 1}College of Computer Science and Electronic Engineering, Hunan University, Changsha, China\\
    \textsuperscript{\rm 2}School of Computer Science and Engineering, Central South University, Changsha, China\\
    \{yanwenba, xuan\_liu, chenxinning, wonhow, xuyangcs, lkl\}@hnu.edu.cn, sgzhang@csu.edu.cn
}
\usepackage{bibentry}
\begin{document}

\maketitle

\begin{abstract}
While decentralized training is attractive in multi-agent reinforcement learning (MARL) for its excellent scalability and robustness, its inherent coordination challenges in collaborative tasks result in numerous interactions for agents to learn good policies.
To alleviate this problem, action advising methods make experienced agents share their knowledge about what to do, while less experienced agents strictly follow the received advice.
However, this method of sharing and utilizing knowledge may hinder the team's exploration of better states, as agents can be unduly influenced by suboptimal or even adverse advice, especially in the early stages of learning.
Inspired by the fact that humans can learn not only from the success but also from the failure of others, this paper proposes a novel knowledge sharing framework called \textit{Cautiously-Optimistic kNowledge Sharing} (CONS). 
CONS enables each agent to share both positive and negative knowledge and cautiously assimilate knowledge from others, thereby enhancing the efficiency of early-stage exploration and the agents' robustness to adverse advice. 
Moreover, considering the continuous improvement of policies, agents value negative knowledge more in the early stages of learning and shift their focus to positive knowledge in the later stages.
Our framework can be easily integrated into existing Q-learning based methods without introducing additional training costs.
We evaluate CONS in several challenging multi-agent tasks and find it 
excels in environments where optimal behavioral patterns are difficult to discover, surpassing the baselines in terms of convergence rate and final performance.
\end{abstract}

\section{Introduction}
Cooperative multi-agent reinforcement learning (MARL) has attracted much attention in recent years due to its ability to solve complex real-world problems, such as multi-robot control~\cite{multiRobot2} and traffic scheduling~\cite{cityflow}.
Most of the currently proposed MARL algorithms follow the paradigm of \textit{centralized training and decentralized execution} (CTDE)~\cite{maddpg, qmix,qtran,facmac}, 
where a centralized critic collects information from all agents during the training phase to learn decentralized agent policies. 
However, this paradigm struggles with the huge joint state-action spaces that grow exponentially with the number of agents, and the ideal conditions for deploying centralized critics are often lacking in reality.
In contrast, the paradigm of \textit{decentralized training and decentralized execution} (DTDE)~\cite{iql,idqn} is more scalable and robust, and more adaptable to harsh real-world conditions, as it does not require a centralized critic. 

While the DTDE paradigm has many advantages, it inevitably faces coordination difficulties in collaborative tasks due to the lack of an explicit centralized coordinator and partial observability. 
Agent teams that follow the DTDE paradigm often need to spend a lot of time exploring to develop good strategies.
To alleviate this problem, some communication-based MARL methods focus on allowing proper exchange of local information about observations among agents while following the DTDE paradigm. 
This information can be regarded as the perceptual-level knowledge, allowing agents to make decisions from a broader perspective.
It is usually fed directly into the receiver's network~\cite{atoc,ic3net,i2c}, which expands the local policy spaces of agents and significantly increases the training burden. Furthermore, high-dimensional information in the network has uncertainty and uninterpretability.

Unlike general communication-based methods, action advising shares policy-level rather than perceptual-level knowledge in a more direct and explainable manner, where more experienced agents advise less experienced agents on the best actions and the suggested actions will be executed directly by the advisees. This scheme speeds up coordination among agents and resembles the common way humans communicate---we often provide helpful advice based on our knowledge and beliefs rather than simply providing our own information.
However, agents may give many suboptimal or even poor suggestions, especially in the early stages of learning. 
These inappropriate suggestions not only waste precious communication budget, but also hinder the team from further exploring better states, diminishing the advantages of action advising in complex environments that require extensive exploration.
Furthermore, the suggestions provided are also one-sided.

To make knowledge sharing more comprehensive and mitigate the negative impacts of inappropriate knowledge, we propose a novel knowledge sharing framework called \textit{\textbf{C}autiously-\textbf{O}ptimistic k\textbf{N}owledge \textbf{S}haring} (CONS)
\footnote{We provide open-source implementations of CONS in https://github.com/byw0919/CONS}.
Specifically, inspired by the fact that humans can learn not only from the success of others but also from their failure, CONS agents share experiences of both failure and success (i.e., negative and positive knowledge) simultaneously, whereas previous works only involve the latter. 
Unlike agents in other action advising methods, CONS agents do not simply adopt the suggested actions after receiving knowledge from others.
Instead, they incorporate the received knowledge by softly updating their action probabilities, thereby forming new policy.
Subsequently, CONS agents conduct targeted exploration based on their new policy and their confidence.
They are optimistic due to the belief that sharing and learning from negative knowledge is beneficial; at the same time, they are cautious due to not blindly following the acquired knowledge.
It should to be emphasized that CONS only affects the action selection of the agent, not the training process of the underlying algorithm.
Therefore, CONS can be easily integrated into existing Q-learning based methods without introducing additional training overhead.
Experimental results show that CONS performs well in environments where the optimal behavioral patterns are harder to discover compared to suboptimal ones, surpassing the baselines in terms of convergence rate and final performance.

\section{Related Work}
\subsection{Decentralized Training and Decentralized Execution (DTDE)}
With excellent scalability and robustness, DTDE is a promising paradigm for using MARL to solve real-world problems.
There is no centralized critic in DTDE paradigm so agents only use local information to make decisions during both training and execution. 
The most straightforward way to use the DTDE paradigm in MARL is to make each agent run a single-agent reinforcement learning algorithm independently~\cite{iql}, such as independent Q-learning (IQL)~\cite{idqn} and independent PPO (IPPO)~\cite{ippo}.
In addition to the aforementioned intuitive algorithms, some works also focus on other aspects of decentralized training.
Hysteretic Q-learning~\cite{hysteretic-ql} and lenient Q-learning~\cite{lenient-ql} let agents be optimistic and appropriately ignore value penalties, thereby promoting team cooperation. 
Ideal independent Q-learning (I2Q)~\cite{i2q} alleviates environmental non-stationarity by having each agent model an ideal transition function and perform independent Q-learning on it.
Please note that this paper focuses on knowledge sharing rather than decentralized algorithms, so we directly use deep recurrent Q-network (DRQN)~\cite{drqn} as an instance of the underlying algorithm to implement CONS.

\subsection{Knowledge Sharing}
Knowledge sharing speeds up learning, fosters coordination among agents, and has various forms.

\paragraph{Communication.}
Communication-based methods usually share local observations (or observation embeddings) of agents.
ATOC~\cite{atoc} agent uses an attention unit to decide whether to communicate or not, and if so, selects several collaborators in its observable field to communicate.
IC3Net~\cite{ic3net} extends CommNet~\cite{commnet} by using the gating mechanism to determine whether to broadcast messages on a common channel.
I2C~\cite{i2c} agent learns prior knowledge for agent-agent communication through causal effect to capture the necessity of communication.
GA-Comm~\cite{ga-comm} employs attention to decide which pair of agents can communicate, thereby learning a shared undirected communication graph. 
However, these methods all expand local policy spaces of agents and make learning more difficult.

\paragraph{Experience Sharing.}
Agents in experience sharing methods like SEAC and SEQL~\cite{seac} acquire the trajectories of others as off-policy data to train their own networks, without increasing learning complexity. 
However, the strong assumption that agents have access to others' private trajectory data and the huge amount of information exchanged make it less attractive.

\paragraph{Advising Mechanism.}
Advising mechanism shares policy-level knowledge, where less experienced agents can take good actions without making decisions themselves.
Unfortunately, many methods based on advising mechanism assume the teacher has a well-trained policy~\cite{AdviceImitation,anand2021enhanced,EAA}, or have a centralized information structure~\cite{LeCTR,HMAT,gupta2021hammer}, or increase training overhead~\cite{AdviceImitation}, or are limited to two agents~\cite{LeCTR,HMAT}. 
Two works that are similar to ours are AdHocTD~\cite{adhoc} and PSAF~\cite{PSAF}, but both are based on tabular Q-learning and lack robustness to suboptimal advice.
Unlike the mentioned knowledge sharing methods, CONS agents modify their action probabilities according to the received policy-level knowledge, and then explore in a targeted manner based on the modified probabilities, avoiding increased training overhead and being robust to suboptimal advice. 
Moreover, CONS agents learn from scratch and can act as both teachers and students during the learning process.

\section{Background}
\subsection{Problem Formulation}
A general cooperative multi-agent reinforcement learning problem can be typically modeled as a partially observable Markov games for $n$ agents~\cite{markovgame}, which is defined by the tuple $(\mathcal{N}, \mathcal{S}, \mathcal{O}, \mathcal{A}, \Omega, \mathcal{P},\{\mathcal{R}^i\}_{i \in \mathcal{N}}, \gamma)$. Here $\mathcal{N}=\{1, \ldots, n\}$ is the set of agents; $\mathcal{S}$ is the state space; $\mathcal{O}=O^1 \times \ldots \times O^n$ is the joint observation space; $\mathcal{A}=A^1 \times \ldots \times A^n$ is the joint action space. 
At each time step, each agent $i \in \mathcal{N}$ can only access local observations $o^i \in O^i$ drawn from the observation function $\Omega (s,i)$ where $s \in \mathcal{S}$, and then choose an action $a^i \in A^i$ according to its policy $\pi_i(a^i|o^i): {O}^i \mapsto {A}^i$. The transition function $\mathcal{P}(s'|s,\bm{a}): \mathcal{S} \times \mathcal{A} \mapsto \Delta (\mathcal {S})$ returns a distribution over successor states $s'$ given state $s$ and joint action $\bm{a}$. 
After that, each agent $i$ receives an individual reward $r_t^i$ based on its reward function $\mathcal{R}^i: \mathcal{S} \times \mathcal{A} \times \mathcal{S} \mapsto \mathds{R}$ at time step $t$. 
The purpose of each agent $i$ is to find a policy $\pi_i$ that maximize its expected discounted return $\mathds{E}[\sum_{t=0}^{H} \gamma^t r_t^i|\pi_i]$ over horizon $H$, where $\gamma$ is the discount factor. 

In this work, we focus on general cooperative tasks that do not require agents to share the same reward at each time step, but share the same behavior patterns that lead to rewards.
Besides, we assume $O^1= \ldots =O^n=O$ and $A^1= \ldots =A^n=A$.
\subsection{Advising Mechanism} 
Advising mechanism is an efficient, scalable knowledge sharing paradigm, where experienced agents (acting as teachers) give advice to less experienced agents (acting as students) on what to do based on their positive knowledge.
CONS follows the teacher-student relationship in the advising mechanism, but innovates in the content of advice and the way of using advice. 
To make our work easy to understand, here we introduce a promising advising framework, AdHocTD~\cite{adhoc}, which focuses on when to ask for and when to give advice within two budgets: $b_{ask}$ and $b_{give}$. 
To encourage agents to engage in these behaviors only in critical states, the probabilities for them to ask for and give advice can be calculated as
\begin{equation}
	P_{ask}(s)={\left(1+\upsilon_a\right)}^{-f(s)}
	\label{eq:1}
\end{equation}
and
\begin{equation}
	P_{give}(s)={1-\left(1+\upsilon_g\right)}^{-g(s)},
	\label{eq:2}
\end{equation}
respectively. 
$f(s)=\sqrt{n_{visit}(s)}$ is the confidence function of asking for advice for the current state $s$, where $n_{visit}(s)$ is the number of times the agent has visited $s$. 
$g(s) = f(s)|\max_{a}Q(s,a)-\min_{a}Q(s,a)|$ is the confidence function of giving advice for the state $s$, where $Q$ is the agent's Q-network and $|{max}_aQ(s,a)-{min}_aQ(s,a)|$ measures the importance of state $s$.
In the above two equations, $\upsilon_a$ and $\upsilon_g$ are pre-defined scaling variables.
When receiving one advice, an AdHocTD agent follows it exactly; when receiving more than one advice, it selects the executed action through a majority vote.

\subsection{Independent Q-learning}
Independent Q-learning (IQL)~\cite{idqn} combines deep Q-network (DQN)~\cite{dqn} with independent learning~\cite{iql}, where each agent runs the DQN algorithm independently.
The Q-function of each agent $i$ that estimates the value of each state-action pair is $Q^\pi(s,a)=\mathds{E}[R|s^t=s, a^t=a]$ (the superscript $i$ is omitted for simplicity), where $\pi$ is its policy, $R$ is its total discounted return, $s$ is the current state and $a$ is the action it chooses.
The optimal Q-function $Q^*(s,a)=\max_{\pi}Q^{\pi}(s,a)$ obeys the Bellman optimality equation $Q^*(s,a)=\mathds{E}_{s'}[r+\gamma \max_{a'}Q^*(s',a')]$.
DQNs are optimized by minimizing
\begin{equation}
	\mathcal{L}(\theta)=\mathds{E}_{(s,a,r,s') \sim \mathcal{D}}[y-Q(s,a;\theta))^2]\label{eq:3},
\end{equation}
where $\mathcal{D}$ is experience replay buffer and $y=r+\gamma \max_{a'}Q(s',a';\bar{\theta})$.
The parameters $\bar{\theta}$ of the agent's target network are periodically copied from $\theta$ and remain constant for a certain number of iterations.
Other variants of DQN, such as DRQN~\cite{drqn}, can also be combined with independent learning while keeping the loss function unchanged in form.

\section{Method}
\begin{figure}[t]
	\centering
	\includegraphics[width=1.0\linewidth]{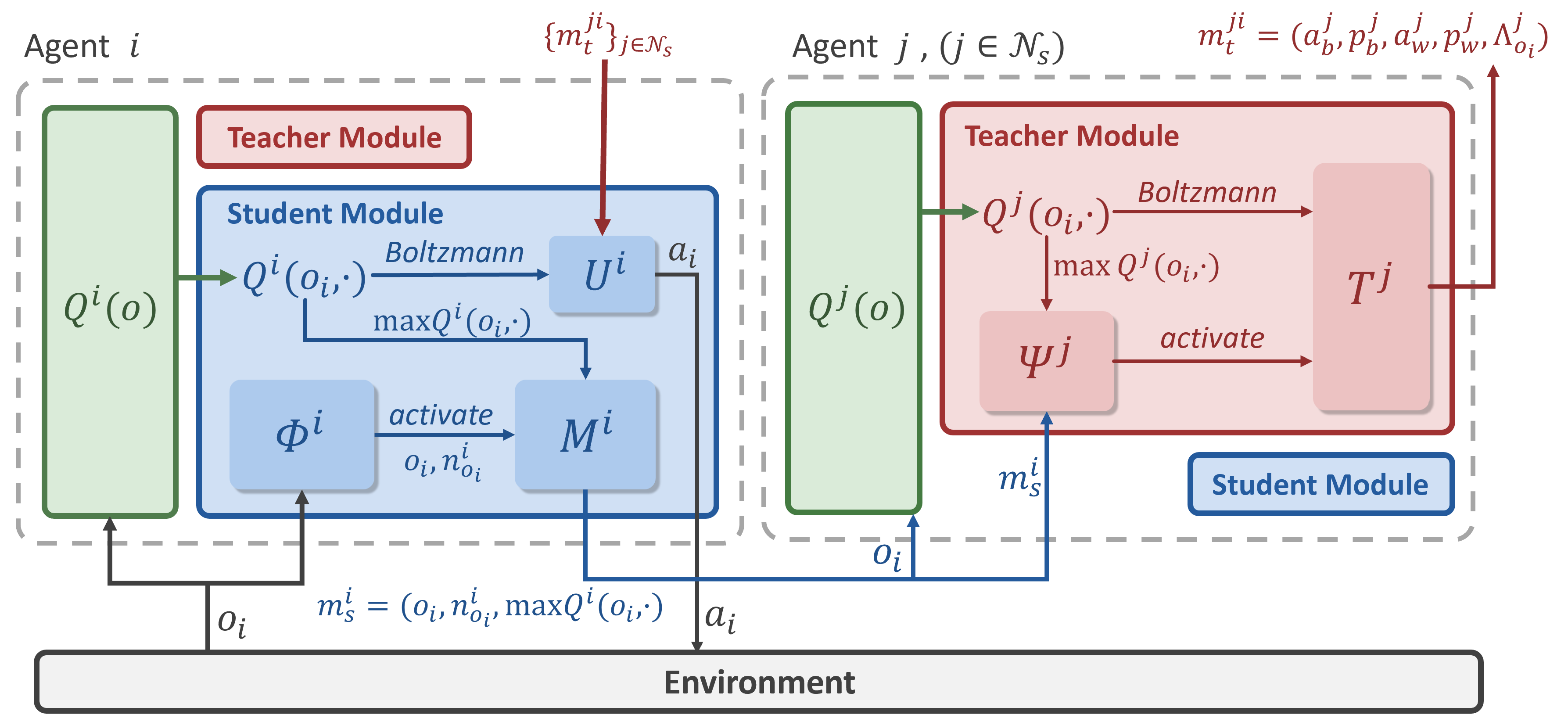}
	\caption{An overview of CONS. 
		\textbf{(Student)} 
		$\mit \Phi^i$ receives $o_i$, sends it with $n_{o_i}^i$ to activate $M^i$ with the probability $P_{ask}(o_i)$. 
		If $M^i$ is activated, it assembles $o_i$, $n_{o_i}^i$ and $\max Q^i(o_i,\cdot)$ into a student request message $m_s^i$ and broadcasts it.
		Upon receiving messages $m_t^{ji}$ from the teachers $j \in \mathcal{N}_s$ ($\mathcal{N}_s$ is the set of agents sharing knowledge), $U^i$ modifies the original $\pi^i(\cdot|o_i)$ that derived from $Q^i(o_i,\cdot)$ according to the messages and then samples an action $a_i$ from the modified policy.
		\textbf{(Teacher)}
		$\mit \Psi^j$ decides whether to share knowledge with agent $i$ according to $m_s^i$, and if so, activates module $T^i$. 
		Module $T^i$ extracts positive knowledge $a_b^j$ and $p_b^j$ as well as negative knowledge $a_w^j$ and $p_w^j$ from its policy $\pi^j(\cdot|o_i)$, and then combines them with its prestige $\Lambda_{o_i}^j$ to form a teacher message $m_t^{ji}$ to reply to agent $i$.
	} 
	\label{fig:CONS}
\end{figure}

In this section, we propose CONS, a novel knowledge sharing framework that leverages two types of knowledge and reduces the negative effects of suboptimal knowledge on agents.
We first introduce policy confidence to quantify the level of certainty of the agent's policy and then provide a detailed description of the three stages of CONS. 
Please note that knowledge sharing is initiated only after agents have interacted with the environment for a short period to avoid ineffective sharing in the very early stages.
Without loss of generality, we assume that in the following, agent $i$ takes the role of student, while the other agents may take the role of teachers (uniformly represented as $j$).
Figure~\ref{fig:CONS} shows how CONS works after sharing is initiated.

\subsection{Policy Confidence}
Policy confidence should assess the certainty of agent policy under certain observation $o$ across different $|A|$. So we define it as the min-max normalized value of the standard deviation of the action probability distribution:
\begin{equation}
	\Gamma_{o} =\frac{\sigma_o-\min(\sigma_o)}{\max(\sigma_o)-0}=\frac{|A|\sigma_o}{\sqrt{|A|-1}}\label{eq:4}.
\end{equation}
In the above equation, $\sigma_o$ denotes the standard deviation of the action probability distribution $\pi(\cdot|o)$ conditioned on observation $o$ and $A$ denotes the agent's action space. $\sigma_o$ reaches its maximum value when only one action has the maximum probability of 1, while its minimum value occurs when the probabilities of all actions are equal to $\frac{1}{|A|}$. $\pi(\cdot|o)$ is derived from the Boltzmann distribution
\begin{equation}
	\pi\left(a \mid o\right)=\frac{e^{Q\left(o, a\right) / T}}{\sum_{k} e^{Q\left(o, a_k\right) / T}}=p_a\label{eq:5},
\end{equation}
where $p_a$ is the probability of action $a$ and $T$ is the temperature parameter used to adjust the randomness of decisions and we set it to 1.

\subsection{Stage 1: Student Sends Request}
After knowledge sharing is initiated, agent $i$ checks its budget $b_{ask}^i$. If not exhausted, it broadcasts a student request message $m_s^i$ with the probability of $P_{ask}(o_i)$ calculated by Eq.~\ref{eq:1}, where $o_i$ is its current observation; otherwise, it samples an action from its own policy.
In addition to $o_i$, $m_s^i$ also includes $n_{o_i}^i$ and $\max Q^i(o_i,\cdot)$, representing the number of times agent $i$ has observed $o_i$ and its corresponding maximum Q-value respectively. 

\subsection{Stage 2: Teacher Shares Knowledge}
\label{teacher_share}
Teachers in CONS share both positive and negative knowledge regarding $o_i$ with the student.
Upon receiving $m_s^i=(o_i, n_{o_i}^i,\max Q^i(o_i,\cdot))$, agent $j$ first checks its budget $b_{give}^j$.
If the budget is exhausted, no response will be provided;
otherwise, agent $j$ decides whether to share policy-level knowledge about $o_i$ with agent $i$ based on $m_s^i$, $n_{o_i}^j$ and $\max Q^j(o_i,\cdot)$.
CONS agents are well-intentioned, aiming to share knowledge only at appropriate times, thereby avoiding any potential misinformation.
Specifically, the module $T^j$ in Figure~\ref{fig:CONS} is activated for knowledge extraction only when agent $j$ has more or better experience compared to agent $i$ with respect to observation $o_i$.
This activation condition, which also helps reduce unnecessary communication overhead, can be expressed as 
\begin{equation}
	\mathds{1}_{n_{o_i}^j>n_{o_i}^i} + \mathds{1}_{\max Q^j(o_i,\cdot)>\max Q^i(o_i,\cdot)}>0,
	\label{eq:6}
\end{equation}
where $\mathds{1}$ is the indicator function.
If this inequality holds, it implies that agent $j$ either has observed $o_i$ more frequently or taken more valuable actions under the observation $o_i$ compared to agent $i$.
If $T^j$ is activated, it extracts the knowledge to be shared from the policy distribution $\pi^j(\cdot|o_i)$ derived from Eq.~\ref{eq:5}. 
Along with decision-related knowledge, agent $j$ also shares its local information with agent $i$ so that agent $i$ can calculate the weights of each responding teacher.
Specifically, the teacher message replied by agent $j$ is $m_t^j=(a_b^j, p_b^j, a_w^j, p_w^j, \Lambda_{o_i}^j)$, where $a_b^j$ and $p_b^j$ represent the best action and its probability, $a_w^j$ and $p_w^j$ represent the worst action and its probability, and $\Lambda_{o_i}^j$ represents the prestige of agent $j$. This observation-specific prestige should reflect agent $j$'s familiarity with $o_i$ and confidence in making decisions under $o_i$, which can be defined as
\begin{equation}
	\Lambda_{o_i}^j=\sqrt{n_{o_i}^j}\times \Gamma_{o_i}^j,
	\label{eq:7}
\end{equation}
where $\Gamma_{o_i}^j$ is the policy confidence of agent $j$ under $o_i$ derived from Eq.~\ref{eq:4}.

\subsection{Stage 3: Student Utilizes the Acquired Knowledge}
CONS agents are optimistic---they believe that the teacher's sharing is well-intentioned, and their knowledge, whether positive or negative, can be beneficial to themselves. 
However, CONS agents are also cautious---they do not blindly trust that the teachers' knowledge is always correct.
Therefore, upon receiving knowledge from teachers, CONS agents carefully adjust their action probabilities and conduct targeted exploration based on their new policies.
This process of absorbing and utilizing knowledge involves several specific details, which we will describe below.

\paragraph{The Changing Weights of Positive Knowledge and Negative Knowledge.}
In challenging tasks, agents initially face failure and gain more success as their policies improve. 
Therefore, negative knowledge is valuable in the early learning period as it helps agents narrow down their exploration space, while positive knowledge becomes more valuable later on as it as it enables agents to accomplish tasks more effectively.
CONS adjusts weights for positive and negative knowledge, denoted as $w_p$ and $w_n$ respectively, increasing the former and decreasing the latter progressively during learning. The sum of the two weights is always equal to 1. Specifically, we use
\begin{equation}
	h(x) = \frac{1}{\frac{1-\rm a}{e_i}\cdot x+\rm a} 
	\label{eq:8}
\end{equation}
to generate $w_n$ for the $x^{\rm th}$ episode, then $w_p$ can be obtained by $w_p=1-w_n$. In the above equation, $e_i$ is the episode when knowledge sharing is initiated and $\rm a$ is an hyperparameter that used to adjust the descent rate of $w_n$. 
After knowledge sharing is initiated, $w_n$ first decreases rapidly from 1, followed by a progressively slower decline. 
We avoid using a linear function because agents should quickly shift their focus to positive knowledge, which aligns with the intuition that negative knowledge is more important than positive knowledge only in the early stages of learning.

\paragraph{Soft Updating of Action Probabilities.}
The CONS agents modify their action probabilities according to the received teachers' knowledge. 
They regard the probabilities within each teacher's knowledge as the update targets for their corresponding actions, and perform multi-objective soft updates while taking into account the weights assigned to positive knowledge, negative knowledge and individual teachers.
Assume that agent $i$ has acquired positive knowledge and negative knowledge about action $a_m$ from teachers in set $\mathcal{N}_{m}^b$ and set $\mathcal{N}_{m}^w$ respectively, that is, $a_m=a_b^k=a_w^l \left(k \in \mathcal{N}_{m}^b, l \in \mathcal{N}_{m}^w\right)$. Then agent $i$ modifies the original probability $p_m^i$ of $a_m$ using the following equation:
\begin{equation}
	\begin{aligned}
		\tilde{p}_m^i&=p_m^i+w_p \sum_{k} w_k \cdot \tau \left(p_b^k-p_m^i\right) \cdot \mathds{1}_{p_b^k>p_m^i} \\&+w_n \sum_{l} w_l \cdot \tau \left(p_w^l-p_m^i\right) \cdot \mathds{1}_{p_w^l<p_m^i}
		\label{eq:9},
	\end{aligned}
\end{equation}
where $\tilde{p}_m^i$ is the modified intermediate probability of action $a_m$ for agent $i$ to be subsequently normalized through softmax. 
$\tau \in (0,1)$ controls the update rate, and the indicator function is used to mask inappropriate modifications. 
$w_k$ and $w_l$ represent the weights of teacher $k$ and $l$ respectively, which are calculated based on the prestige $\Lambda^k$ and $\Lambda^l$ (the subscript $o_i$ is omitted for simplicity) using the following equation:
\begin{equation}
	w_k=\frac{e^{\Lambda^k}}{\sum_{k}{e^{\Lambda^k}}},w_l=\frac{e^{\Lambda^l}}{\sum_{k}{e^{\Lambda^l}}}\label{eq:10}.
\end{equation}
The probabilities of other actions in $A$ that are considered best or worst by teachers can be modified in the same way. 

\paragraph{Sample An Action.}
After completing all necessary probability modifications, agent $i$ obtains a new policy $\tilde{\pi}^i(\cdot|o_i)$ by performing softmax normalization on all probabilities, and then calculates its new policy confidence $\tilde{\Gamma}$ (the superscript $i$ and the subscript $o_i$ are omitted for simplicity).
The policy $\tilde{\pi}^i(\cdot|o_i)$ is derived by cautiously absorbing the knowledge from all teachers, which integrates their experiences.
Based on $\tilde{\pi}^i$ and $\tilde{\Gamma}$, agent $i$ performs targeted exploration to sample an action $a_i$ to be executed as algorithm~\ref{alg:1} shows.

\begin{algorithm}[htbp]  
	\caption{Sample an action $a_i$ to be executed through targeted exploration.}  
	\label{alg:1}  
	\begin{algorithmic}[1]
		\Require The new policy $\tilde{\pi}^i(\cdot|o_i)$ and its confidence $\tilde{\Gamma}$.
		\With {probability $\tilde{\Gamma}$} 
		\State $a_i \gets \mathop{\arg\max}\limits_{a} \tilde{\pi}^i(a|o_i)$
		\algorithmiccomment{Sample the best action}
		\EndWith
		\With {probability $1-\tilde{\Gamma}$} 
		\State Divide $[0,1]$ into $|A|-1$ equal intervals
		\If {$\tilde{\Gamma}$ is in the $q^{\rm th}$ interval in ascending order}
		\State Remove the worst $q$ actions from $\tilde{\pi}^i$.
		\Statex \algorithmiccomment{$q\in{1,2,\ldots,(|A|-1)}$}
		\State Normalize the remaining action probabilities to 
		\Statex \qquad \quad policy $\Pi$ to be sampled.
		\State $a_i \gets sample(\Pi)$
		\EndIf
		\EndWith
		\Return $a_i$
	\end{algorithmic}
\end{algorithm}
Why do CONS agents explore after excluding several actions? (i) Exploration (rather than taking the best action) is to gain a comprehensive understanding of the task and avoid getting stuck in a suboptimal solution; (ii) Excluding some low-probability actions can improve their exploration efficiency. 
In addition, the way CONS agents choose actions also conforms the following intuitions. 
A small value of $\tilde{\Gamma}$ indicates that the probabilities of each action are similar, thus agent $i$ should prioritize exploration. Interestingly, at this juncture, there is a low probability of directly sampling the best action, and 
an action will be sampled from a larger subset of $A$.
A large value of $\tilde{\Gamma}$ indicates the exact opposite, i.e., agent $i$ has a high probability of taking the best action directly, and its exploration is more limited.

\begin{table*}[t]
	\centering
	\caption{Two settings for Patient Gold Miner (PGM) environment}
	\label{table:pgm settings}
	\begin{tabular}{ccccccccccc}
		\toprule
		Exp name & Grid size & Agent view & $N$ & $N_g$ & $T_d$ & $R_g$ & $N_p$ & $T_s$ & $L$ & $T_d/L$ \\
		\midrule
		PGM-6ag \textbf{(easier)} & 12x12 & 5x5 & 6 & 2 & 10 & 30 & 3 & 10 & 50 & 20$\%$ \\
		PGM-3ag \textbf{(harder)} & 8x9  & 3x5 & 3 & 1 & 8 & 20 & 2 & 8 & 25 & 32$\%$ \\
		\bottomrule
	\end{tabular}
\end{table*}

\section{Experiments}

In this section, we evaluate the effectiveness of CONS in three cooperative multi-agent tasks: \textit{patient gold miner}, \textit{find the treasure} and \textit{cleanup}. 
Additionally, we study ablations to further demonstrate the significance of negative knowledge sharing, cautious absorption of knowledge and targeted exploration on the team learning efficiency.
Lastly, we discuss the limitations of CONS.
We mainly compare CONS with I2Q~\cite{i2q}, SEQL~\cite{seac}, GA-Comm~\cite{ga-comm}, AdHocTD~\cite{adhoc} and IQL~\cite{idqn}. The diversity exploration method MAVEN~\cite{maven} is also included in the evaluation of the \textit{find the treasure} task, where agents receive global rewards. For all experiments, unless otherwise stated, we run 10 evaluation episodes without any sharing or exploration every 10k episodes. More training details and environment settings can be found in Appendix.

\begin{figure}[htbp]
	\centering
	\subfigure[]{\includegraphics[width=0.255\linewidth]{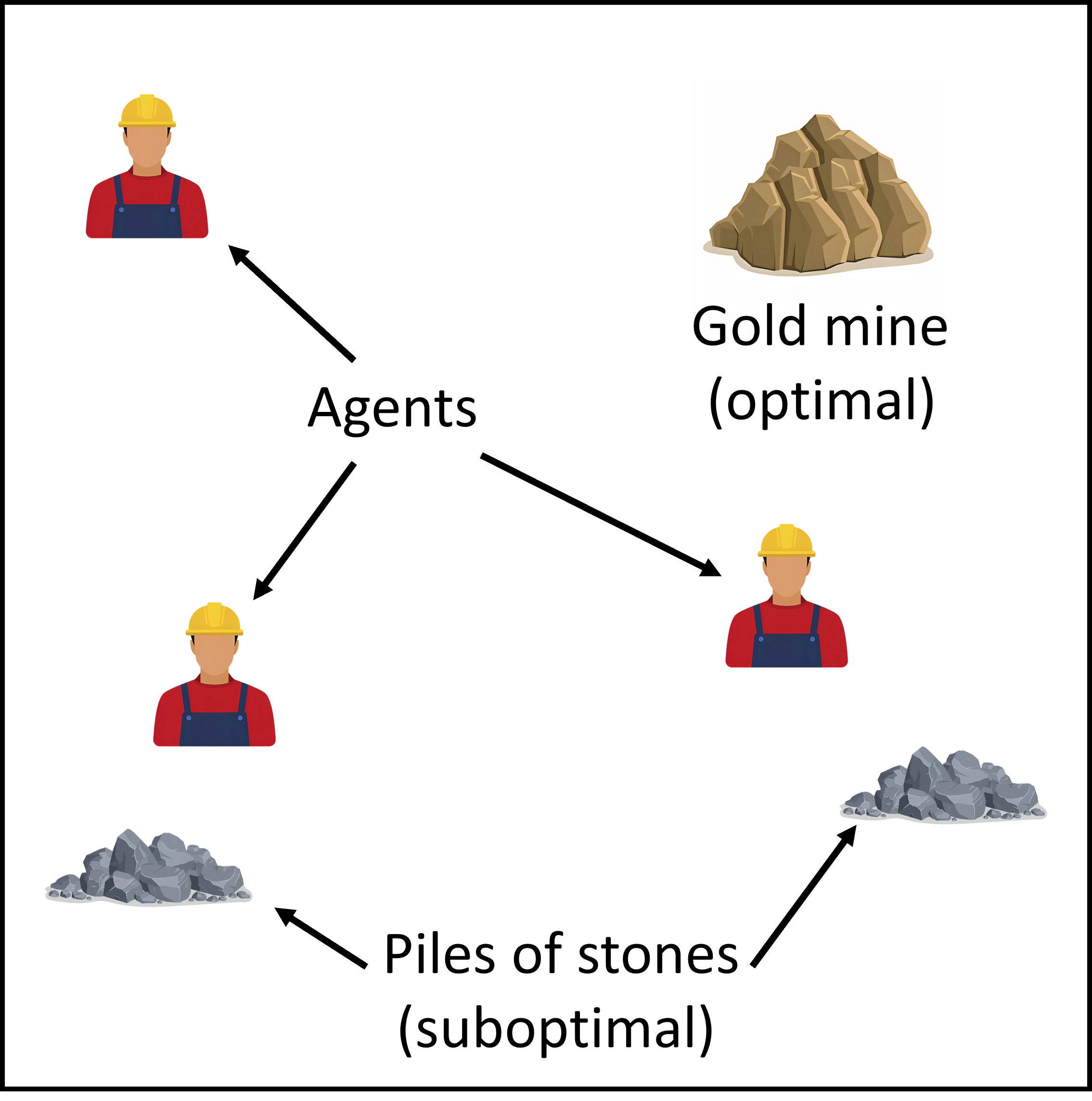}}
	\hspace{3.5mm}
	\subfigure[]{\includegraphics[width=0.26\linewidth]{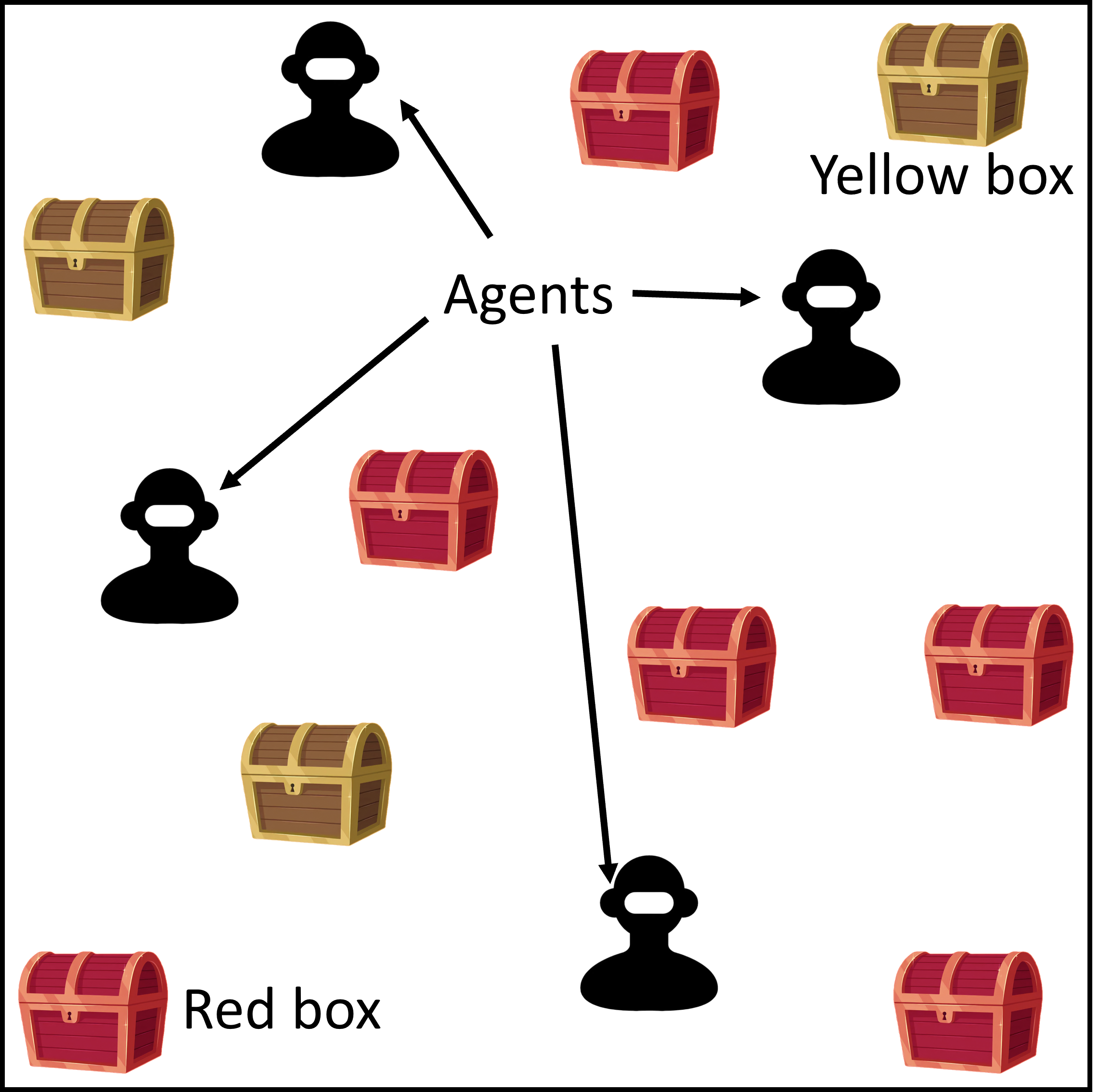}}
	\subfigure[]{\includegraphics[width=0.41\linewidth]{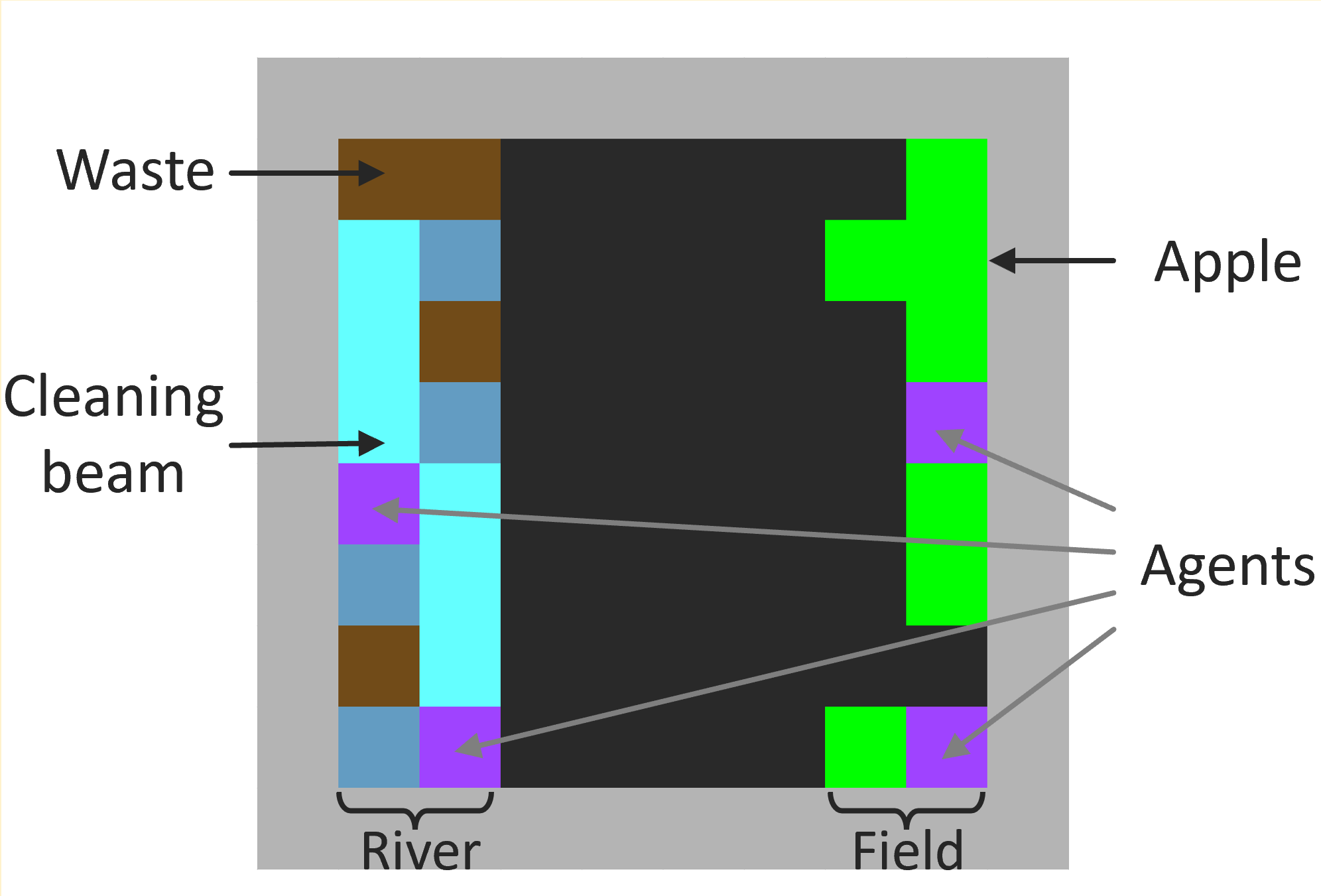}}
	\caption{Illustrations of three environments. (a) Patient gold miner (PGM). (b) Find the treasure (FT). (c) Cleanup.}
	\label{fig:env-figure}
\end{figure}

\subsection{Patient Gold Miner}

\begin{figure*}[t]
	\centering
	\subfigure[]{\includegraphics[width=0.245\linewidth]{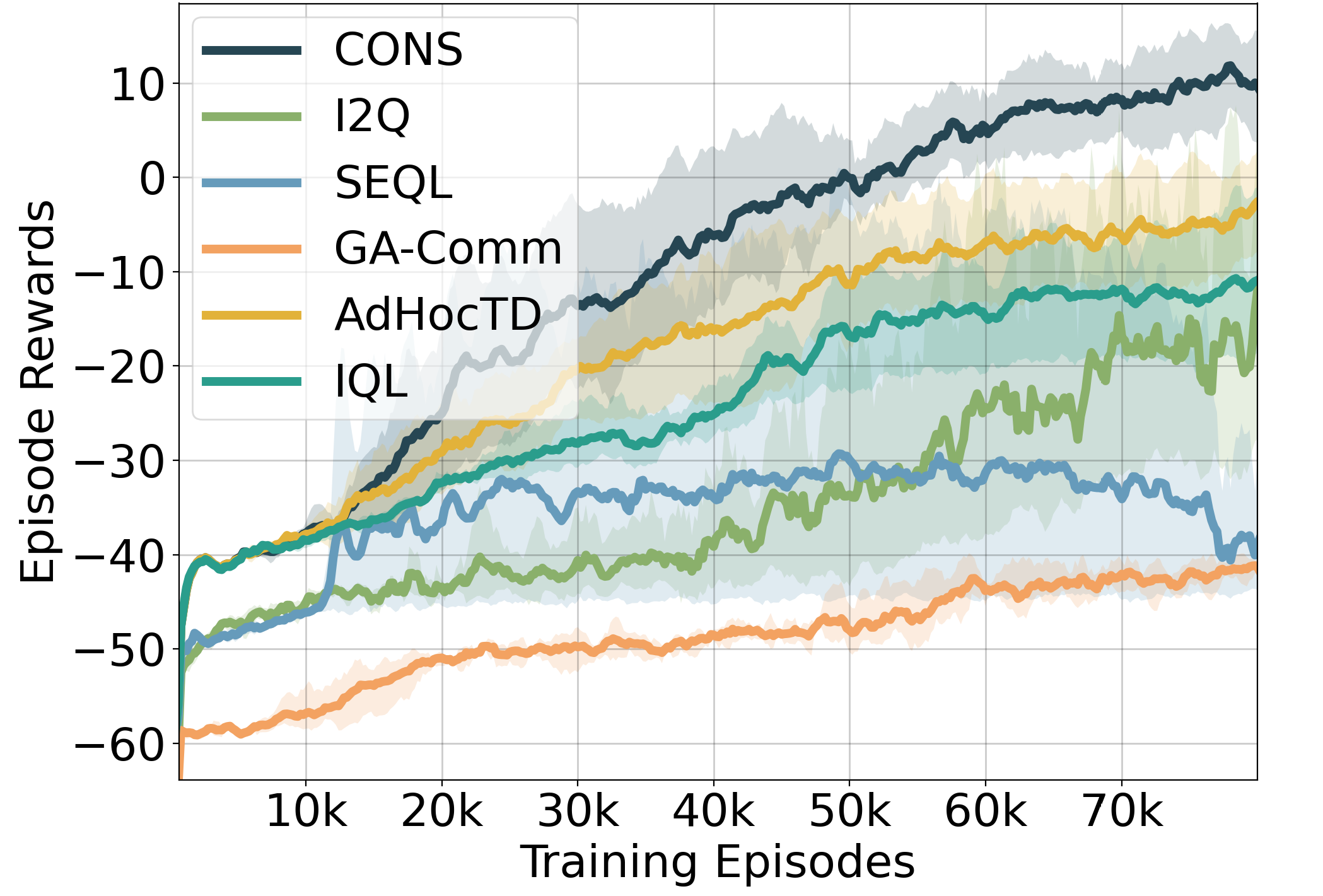}}
	\subfigure[]{\includegraphics[width=0.245\linewidth]{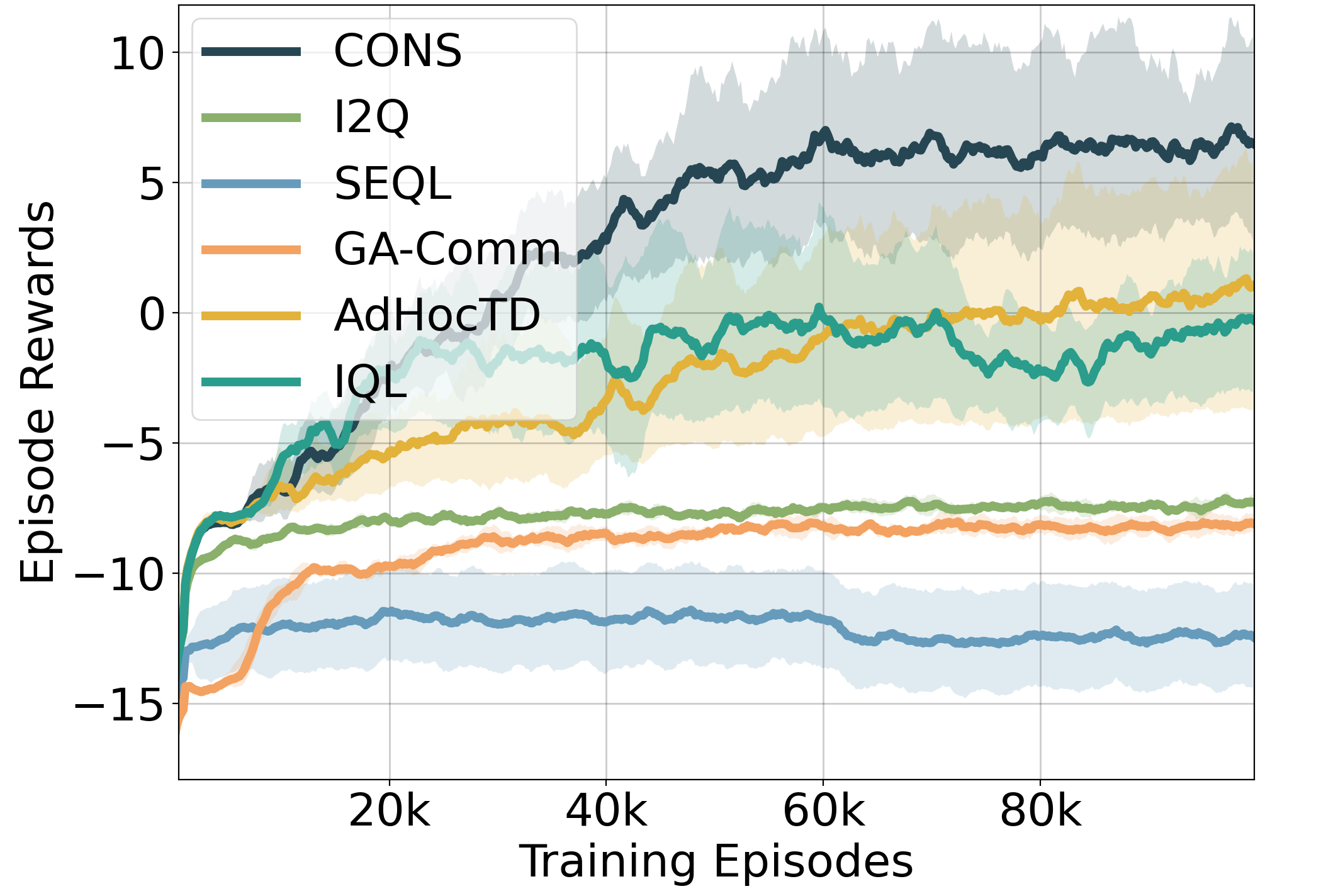}}
	\subfigure[]{\includegraphics[width=0.245\linewidth]{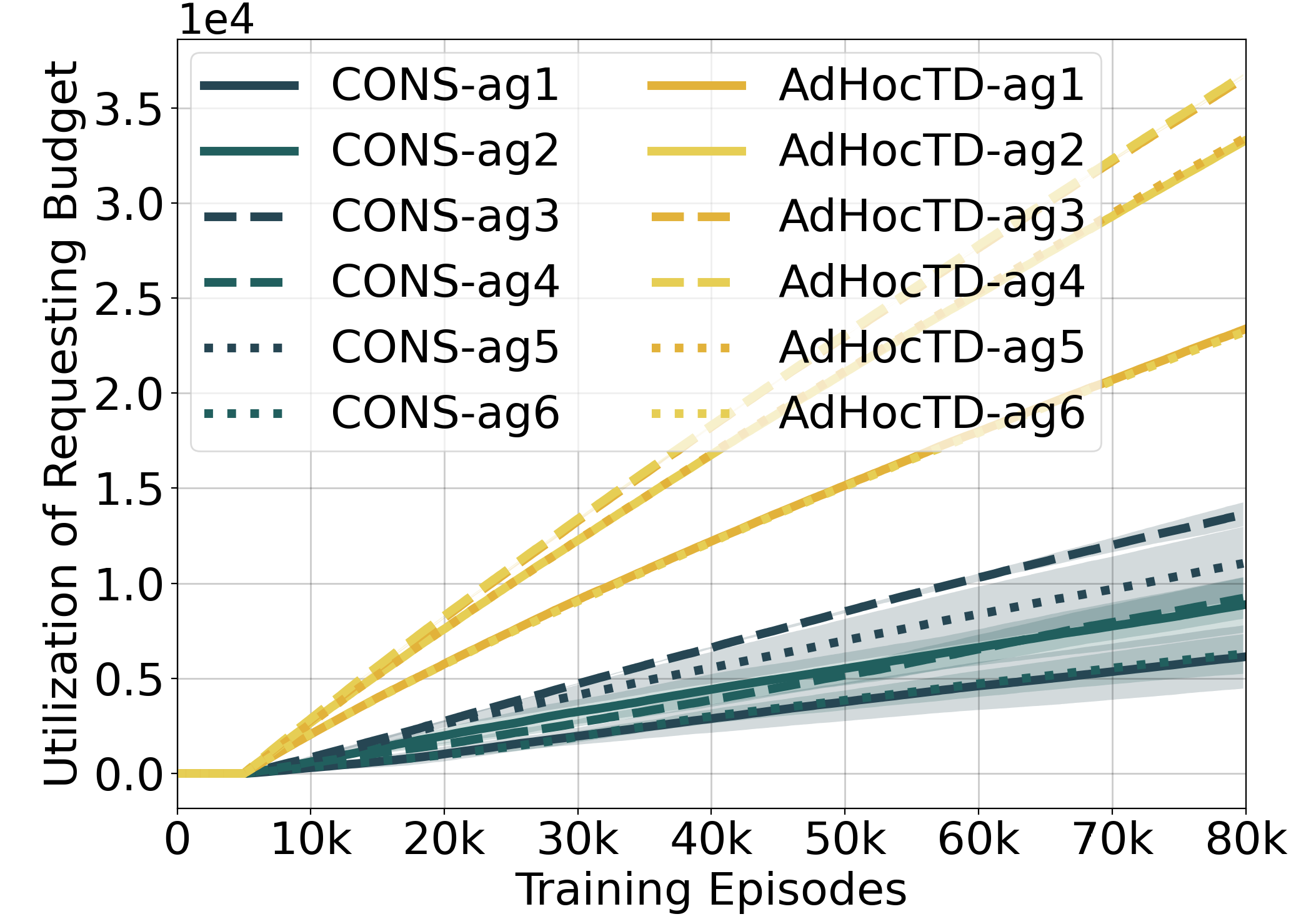}}
	\subfigure[]{\includegraphics[width=0.245\linewidth]{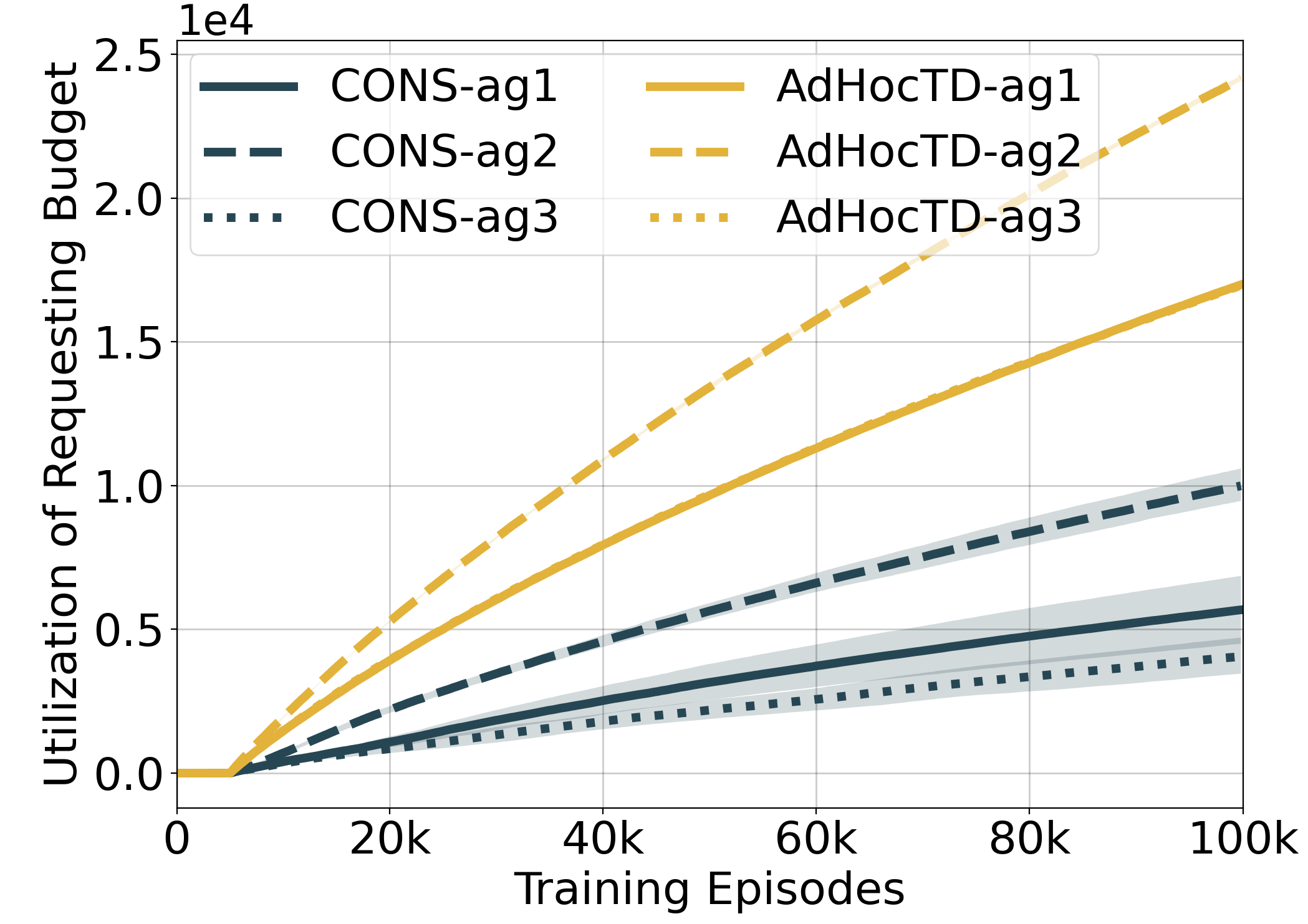}}
	\caption{Experimental results of PGM task. (a-b) Mean evaluating episode returns for the whole team across 5 seeds on PGM-6ag and PGM-3ag with the $95\%$ confidence interval shaded. (c-d) The utilization of requesting budget by CONS agents and AdHocTD agents in PGM-6ag and PGM-3ag.}
	\label{fig:pgm-results}
\end{figure*}

\paragraph{Task Settings.} 
In the patient gold miner (PGM) environment, depicted in Figure~\ref{fig:env-figure}(a), a group of $n$ agents act as miners aiming to maximize their gold collection.
To obtain a piece of gold and receive an individual reward of $r_g$, an agent must spend $T_d$ time steps at a gold mine. However, each step incurs an individual penalty of -1. 
In addition to the $N_g$ gold mines, agents can also get rewards from $N_p$ stone piles without any additional penalties. Each agent can gather one stone per step and receive an individual reward of 0.3. They can obtain a maximum of $T_s$ stones from a stone pile and one piece of gold from a gold mine.
Collecting stones is an easy-to-learn suboptimal behavior, while mining gold is an optimal yet highly risky behavior due to rewards being deeply hidden behind penalties.
We conduct experiments under two different settings, as detailed in Table~\ref{table:pgm settings}. 
Note that the task difficulty is determined by the ratio of $T_d$ to the episode length $L$, not by $n$.
A larger ratio significantly reduces the probability of agents finding the optimal strategy, making PGM-3ag more challenging than PGM-6ag.

\paragraph{Results.} 
Figure~\ref{fig:pgm-results}(a) shows that CONS outperforms other baselines both in sample efficiency and final performance.
It achieves performance equivalent to AdHocTD and IQL in only half the number of episodes and eventually surpasses them.
The IQL agents must independently explore the environment to discern the higher value of gold mines.
For AdHocTD, the preference of experienced agents towards suboptimal behavior is propagated throughout the entire team, leading to a suboptimal outcome.
I2Q spends a lot of time learning the ideal transition model in the early stages, resulting in slower performance improvement.
SEQL agents hardly benefit from others' trajectories due to the rarity of high-value trajectories in this task. Instead, they are overly influenced by numerous suboptimal trajectories, making the discovery of high-value states even more difficult.
Despite the provision of richer information for agents' decision-making, the communication messages in GA-Comm have limited impact on facilitating the transition from suboptimal behavioral patterns to optimal ones. 
Compared to the methods above, CONS agents cautiously assimilate the acquired positive and negative knowledge to update their policy and conduct targeted exploration, thereby achieving a higher learning rate and better performance.
Figure~\ref{fig:pgm-results}(b) shows that CONS still outperforms other baselines on this harder task, and its advantage is more pronounced.
All baselines exhibit a relative performance decrease, indicating that the aforementioned issue worsens with an increase in environmental difficulty.
Figure~\ref{fig:pgm-results}(c) and (d) depict the utilization of requesting budget $b_{ask}$ by CONS and AdHocTD agents throughout the entire learning process. 
In both settings, CONS agents consistently exhibit significantly lower average budget utilization compared to AdHocTD agents, at $30.1\%$ and $33.5\%$ of the budget used by AdHocTD agents, respectively.
This is primarily because CONS avoids many inappropriate knowledge sharing through Eq.~\ref{eq:6}.
The learning curves and budget utilization indicate that CONS can achieve better performance with less knowledge sharing.

\subsection{Find the Treasure}
\begin{figure}[htbp]
	\centering
	\hspace{-3mm}
	\subfigure[]{\includegraphics[width=0.48\linewidth]{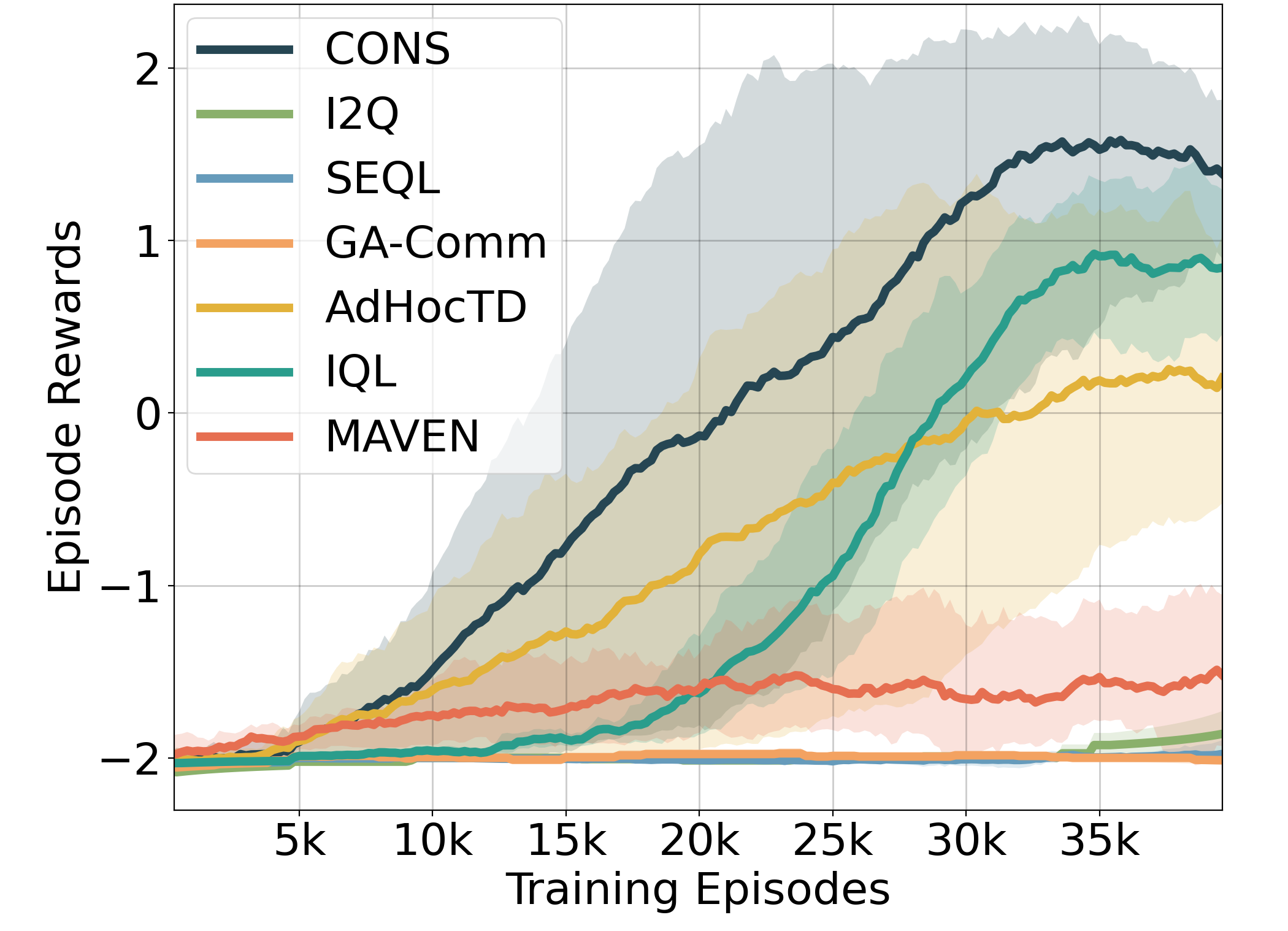}}
	\hspace{1mm}
	\subfigure[]{\includegraphics[width=0.48\linewidth]{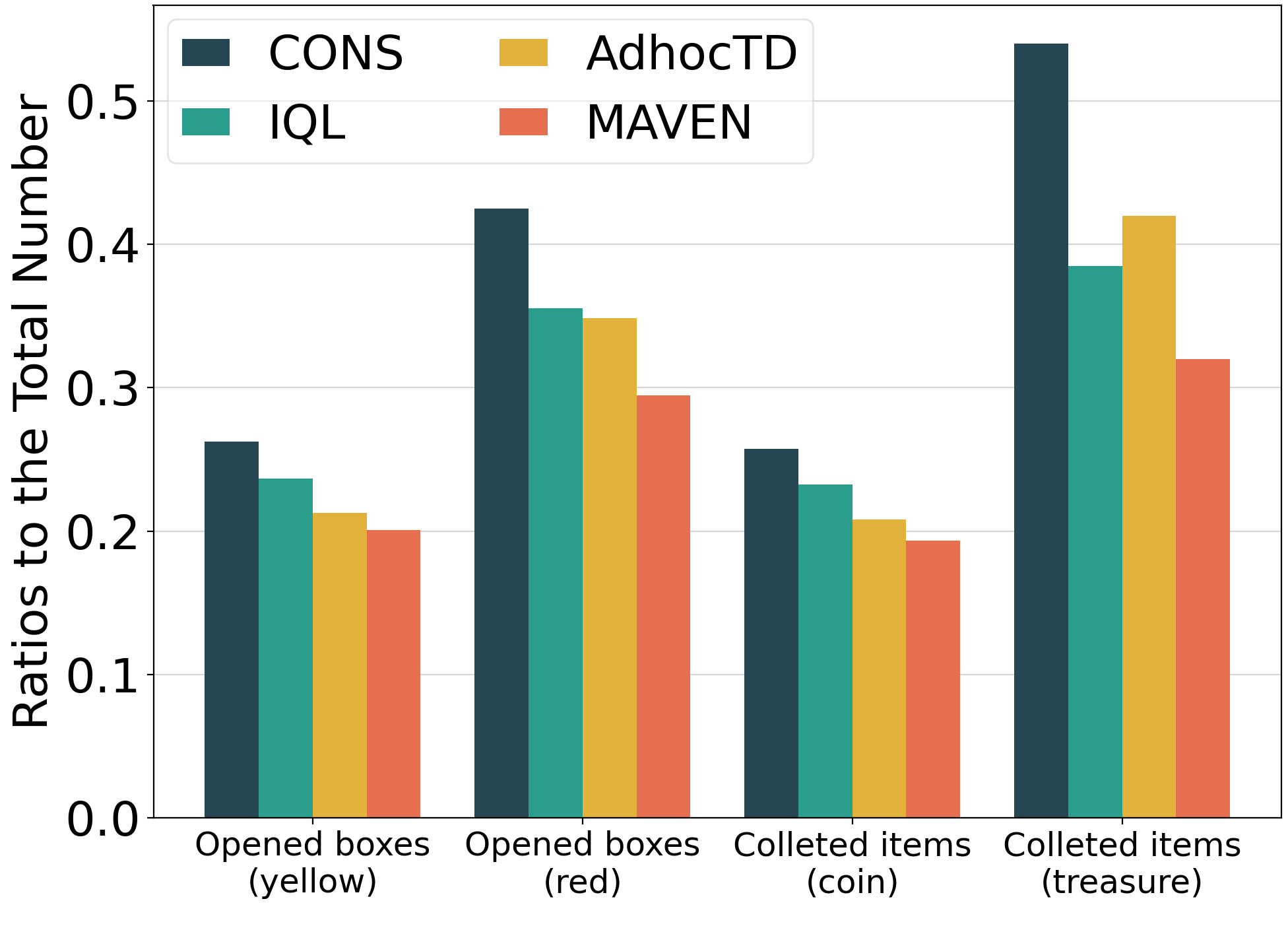}}
	\caption{FT task: (a) Mean evaluating episode returns for the whole team across 4 seeds with the $95\%$ confidence interval shaded. (b) The ratios of opened boxes and collected items to the total number of corresponding entities within 400 evaluation episodes.}
	\label{fig:FT}
\end{figure}

\paragraph{Task Settings.}
In find the treasure (FT) environment depicted in Figure~\ref{fig:env-figure}(b), 4 agents must collaborate to search for a single treasure hidden within one of the 6 red boxes. Collecting this treasure yields a team reward of +15. There are also 3 yellow boxes in this environment, each containing a coin that brings a team reward of +2. Each agent has an action space of [up, down, right, left, open, pick up, stay].To open a box, both agents must perform the \textit{open} action on it simultaneously. The items inside the box can only be collected when at least one agent performs the \textit{pick up} action at the opened box. Opening a yellow box incurs a team cost of -1, while opening a red box incurs a team cost of -2.

\paragraph{Results.}
As shown in Figure~\ref{fig:FT}(a), the I2Q, SEQL and GA-Comm agents completely failed in this sparse reward task.
The diversity exploration method MAVEN also has unsatisfactory performance, suggesting that blind exploration can be counterproductive.
IQL and AdHocTD perform relatively well, surpassing MAVEN. However, CONS learns faster and achieves higher rewards compared to them. CONS agents exhibit better coordination than IQL agents due to knowledge sharing, and demonstrate greater robustness to suboptimal knowledge compared to AdHocTD agents due to the cautious absorption and rational utilization of knowledge.
We run 400 evaluation episodes for each algorithm after training and record the counts of opened boxes (red and yellow) and collected items (coins and treasures).
Figure~\ref{fig:FT}(b) shows the ratios of opened boxes and collected items to the total number of corresponding entities for CONS and baselines (I2Q and SEQL are omitted due to poor performance).
The CONS agents have the best grasp of the environment---they effectively balance the objectives of opening red boxes for treasure search and yellow boxes for coin collection, thus achieving excellent performance.

\subsection{Cleanup}
\label{cleanup}
\paragraph{Task Settings.} 
Cleanup~\cite{lio} is a classic public goods game where agents can earn rewards by collecting apples whose growth rate is negatively correlated with the amount of waste in the river. 
Waste is generated uniformly in the river with a probability of $0.5$ per time step until $40\%$ of the river is covered, at which point apples will not grow either. 
All agents can fire the cleaning beam, which can clean the waste within three cells above the agent.
Figure \ref{fig:env-figure}(c) illustrates the cleanup task where 4 agents collaborate to collect apples in an $8 \times 8$ grid world. 
Obviously, the agents in this task need a well-coordinated division of labor so that the team can obatain more rewards. 

\paragraph{Results.}

\begin{figure}[htbp]
	\centering
	\hspace{-3mm}
	\subfigure[]{\includegraphics[width=0.48\linewidth]{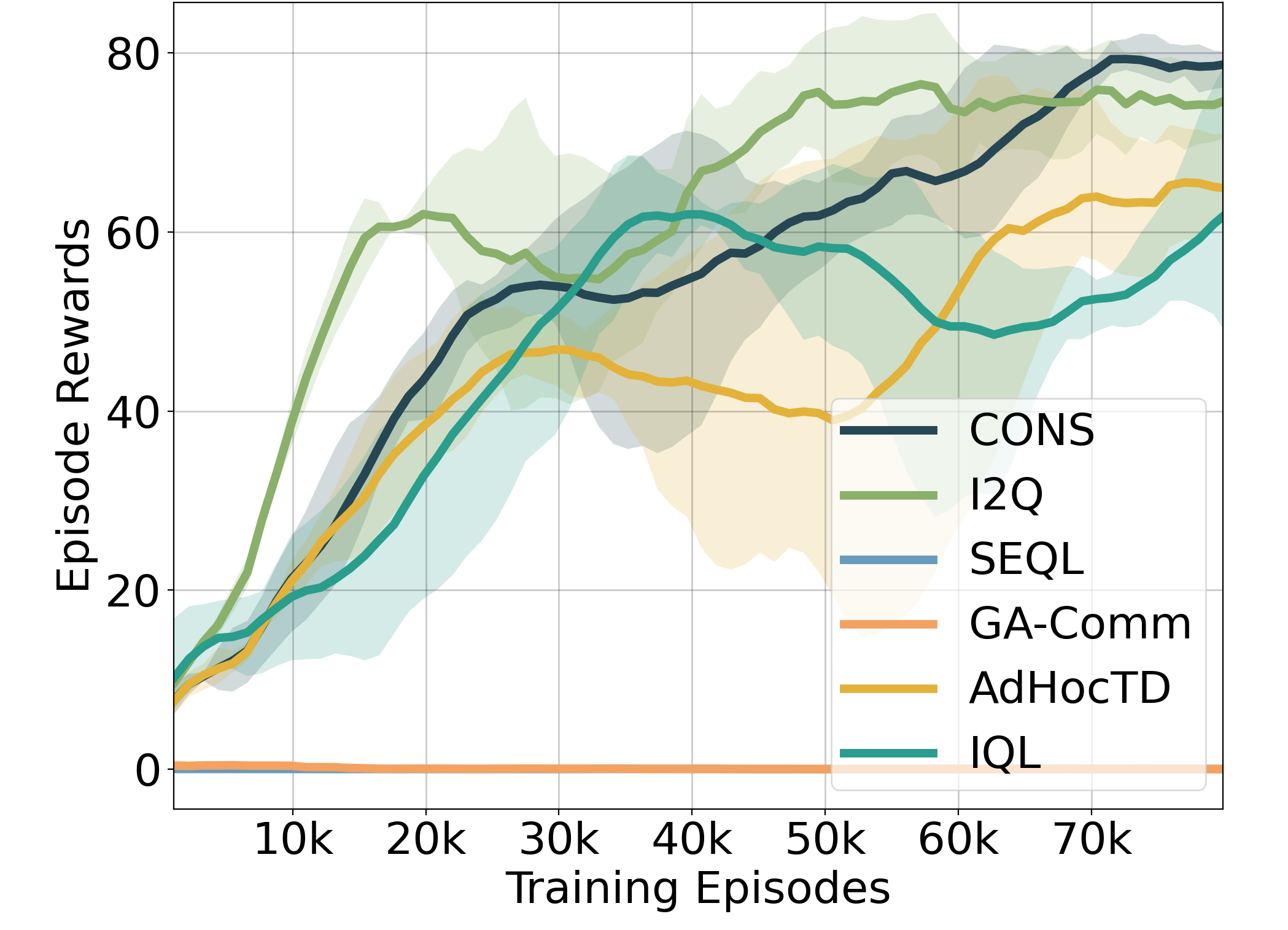}}
	\hspace{1mm}
	\subfigure[]{\includegraphics[width=0.48\linewidth]{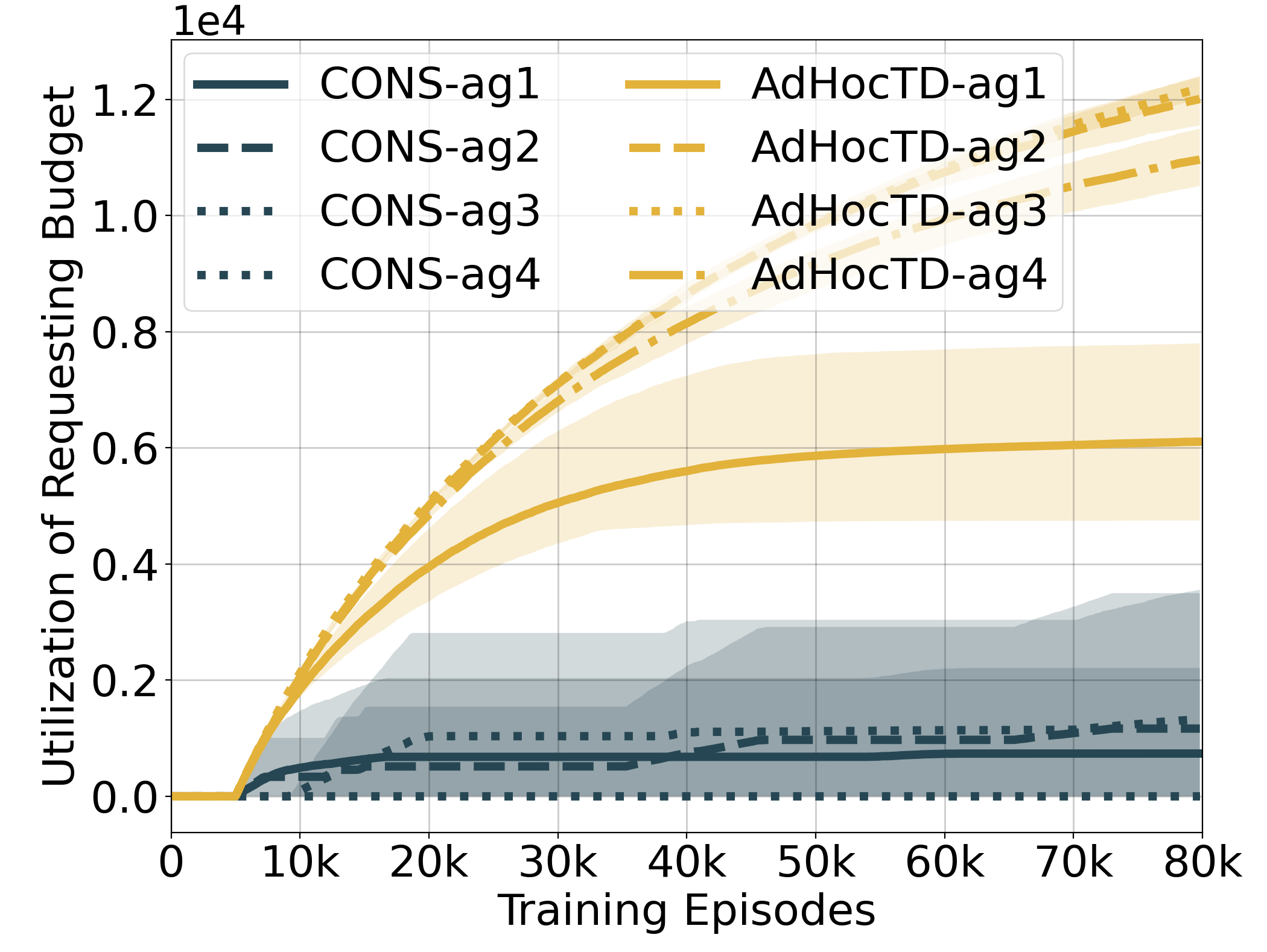}}
	\caption{(a) Mean evaluating episode returns for the whole team across 3 seeds with the $95\%$ confidence interval shaded. (b) The utilization of requesting budget by CONS agents and AdHocTD agents.}
	\label{fig:cleanup}
	
\end{figure}
Cleanup without suboptimal interferences is a simpler task compared to the previous two. However, GA-Comm and SEQL fail completely, as shown in Figure \ref{fig:cleanup}.
Despite achieving acceptable final convergence results, both IQL and AdHocTD exhibit significant performance decline during the middle stage of learning.
This decline can be attributed to agents oscillating between the roles of cleaner and collector, leading to inefficient waste cleaning.
Notably, AdHocTD shows a more prominent performance drop as action advice worsens this oscillation.
Compared with I2Q that performs greedy experience exploitation, the mechanisms in CONS may slightly slow down the learning speed in the early stage. However, CONS efficiently utilizes shared knowledge to fully explore the environment and eventually outperforms I2Q.
In a word, CONS benefits from sharing two types of knowledge and avoids the drawbacks of traditional advising mechanism, resulting in good performance.
Furthermore, Figure ~\ref{fig:cleanup}(b) shows that CONS agents only use $8\%$ of the requesting budget used by AdHocTD agents, again demonstrating the superior performance of CONS.

\subsection{Discussions}
\begin{figure}[htbp]
	\centering
	\hspace{0mm}
	\includegraphics[width=0.7\linewidth]{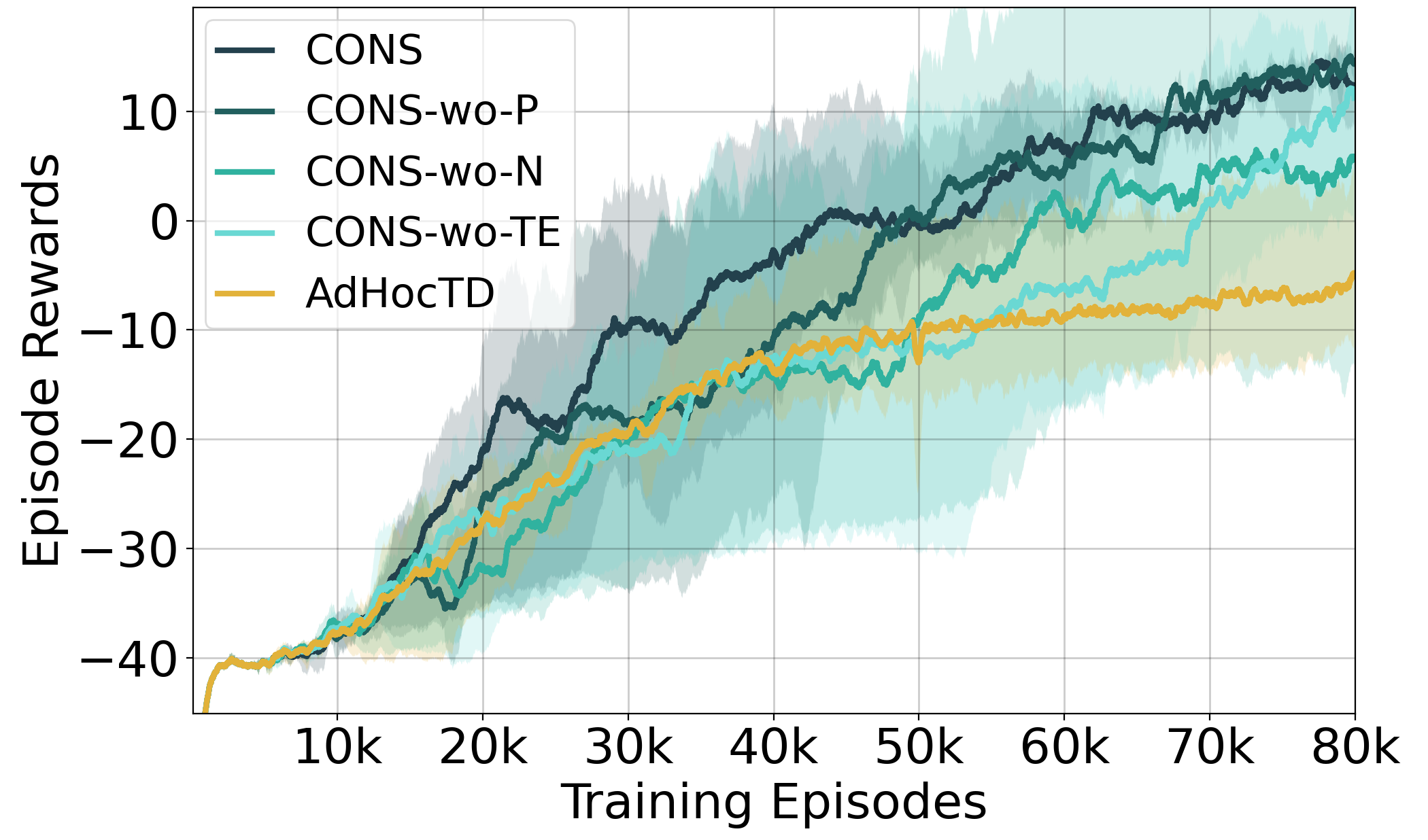}
	\caption{Ablation studies on PGM-6ag. Mean team training rewards are plotted with the $95\%$ confidence interval shaded.}
	\label{fig:ablations}
\end{figure}
\paragraph{Ablations.} To understand the outstanding performance of CONS, we conduct ablation studies on PGM-6ag to evaluate the contribution of its key innovations: negative knowledge sharing, cautious knowledge absorption and targeted exploration. 
We denote CONS without negative knowledge sharing as CONS-wo-N, CONS without positive knowledge sharing as CONS-wo-P, and CONS without targeted exploration (i.e., agents sample actions directly from the modified probability distribution) as CONS-wo-TE. Additionally, the original CONS and AdHocTD are also included in the ablation study.
As shown in Figure~\ref{fig:ablations}, CONS-wo-P outperforms CONS-wo-N, and CONS outperforms CONS-wo-P, indicating that negative knowledge may be more crucial than positive knowledge in challenging tasks, but the presence of positive knowledge can further enhance performance.
Additionally, CONS-wo-TE performs similarly to AdHocTD in the early stages of learning but eventually surpasses it significantly. This indicates that cautious absorption of others' knowledge indeed enhances the agents' robustness to suboptimal advice, thus avoiding falling into suboptimal solutions.
The ablation study results show the effectiveness of negative knowledge sharing, cautious knowledge absorption and targeted exploration.
\paragraph{Limitations.}
CONS is ideal for complex tasks that require extensive exploration or where agents are prone to getting trapped in suboptimal solutions. 
However, when the optimal strategy is evident or no suboptimal solutions exist, strictly following advice would be better as the targeted exploration in CONS may slightly slow down the learning process.
In addition, CONS currently relies on observation counters and value-based underlying algorithms, making it suitable only for discrete tasks for now. 

\section{Conclusion and Future Work} 
In this paper, we propose CONS to maximize the benefits of knowledge sharing for agents. 
The CONS agents share both positive and negative knowledge optimistically and absorb others' knowledge cautiously. 
Experimental results show that CONS can significantly improve learning speed and final performance in challenging tasks. 
For future work, we will extend the underlying idea of CONS to continuous tasks by designing additional networks to replace counters and utilizing action discretization technologies. 

\appendix
\section{Acknowledgments}
This work was supported in part by the National Key Research and Development Program of China under Grant 2022YFC3400404, the National Natural Science Foundation of China under Grant 62172154, 62372473, 62321003, 62002113 and 62272154, the Hunan Provincial Key Research and Development Program of China under Grant 2022GK2004, the Hunan Provincial Natural Science Foundation of China under Grant 2023JJ30702. The authors are grateful for resources from the High Performance Computing Center of Central South University. Prof. Xuan Liu is the corresponding author of the paper.

\bibliography{aaai24}

\newpage
\appendix

\setcounter{secnumdepth}{2}	

\section{Environment Details}
\subsection{Patient Gold Miner}
\begin{figure}[htbp]
	\centering
	\hspace{-1mm}
	\subfigure[PGM-3ag]{\includegraphics[width=0.40\linewidth]{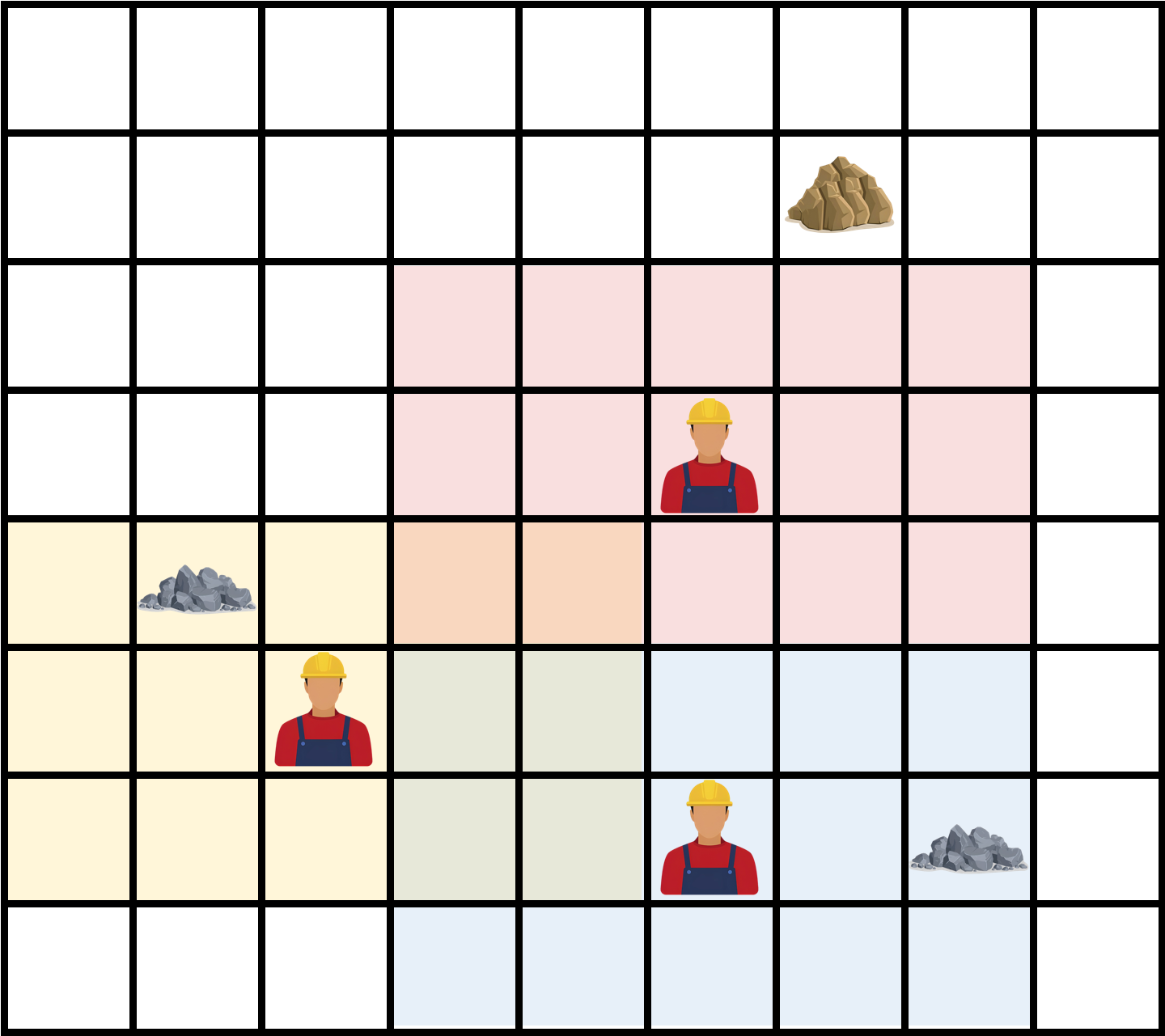}}
	\hspace{3mm}
	\subfigure[PGM-6ag]{\includegraphics[width=0.36\linewidth]{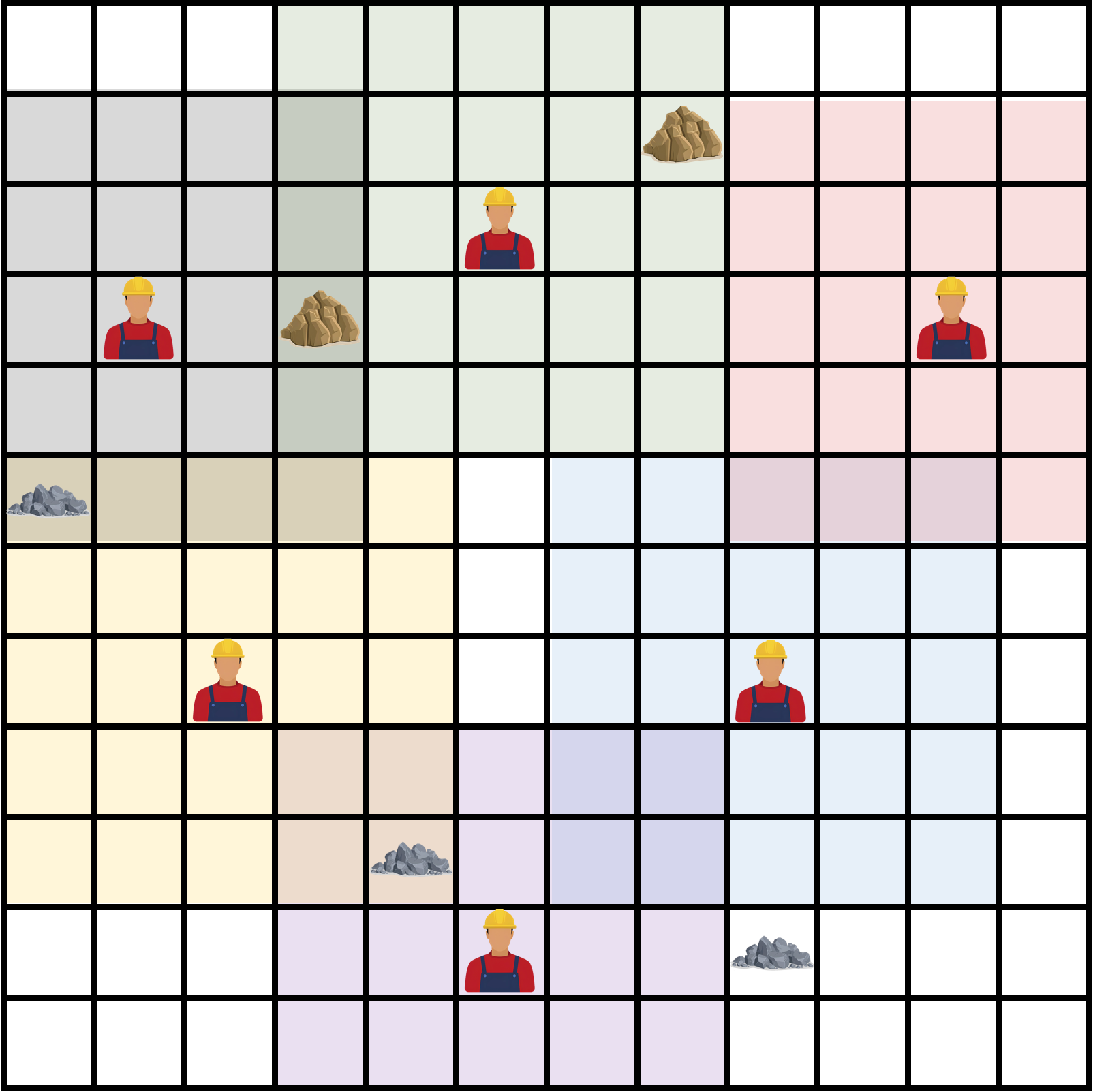}}
	\caption{Two settings of patient gold miner Environment.}
	\label{ap-fig:pgm}
\end{figure}

Patient Gold Miner (PGM) is a complex, partial observable grid world game, which are modified from \textit{ma-gym} environment. In PGM, $n$ agents assume the role of miners, aiming to maximize their gold collection. The action space of each agent is [up, down, right, left, stay]. 
The observation of each agent includes: (i) its own position, (ii) the relative positions of other agents, gold mines and stone piles within its rectangular vision field (as indicated by the filled colors in Figure~\ref{ap-fig:pgm}), and (iii) the current time step. 
When the agent is in the same cell as the gold mine or stone pile, it is considered as interacting with the corresponding resource, including mining the gold mine and collecting a piece of gold or a stone. 
Therefore, Each agent needs to stay in the same grid as the gold mine for at least $T_d$ time steps to acquire a piece of gold.
The possible \textbf{individual} rewards and penalties for this task are shown in Table~\ref{tab:pgm_reward}.

\begin{table}[htbp]
	\centering
	\caption{Possible rewards and penalties \textbf{(individual)} in PGM task.}
	\begin{tabular}{lcc}
		\toprule
		\multicolumn{1}{c}{\multirow{2}[4]{*}{Event}} & \multicolumn{2}{c}{Reward/Penalty} \\
		\cmidrule{2-3}          & \multicolumn{1}{l}{PGM-6ag} & \multicolumn{1}{l}{PGM-3ag} \\
		\midrule
		Collecting a stone & 0.3   & 0.3 \\
		Collecting a piece of gold & 30    & 20 \\
		Mining for gold once & -1    & -1 \\
		Step cost & -0.2  & -0.2 \\
		\bottomrule
	\end{tabular}
	\label{tab:pgm_reward}
\end{table}

\subsection{Find the Treasure}
\begin{figure}[htbp]
	\centering
	\includegraphics[width=0.5\linewidth]{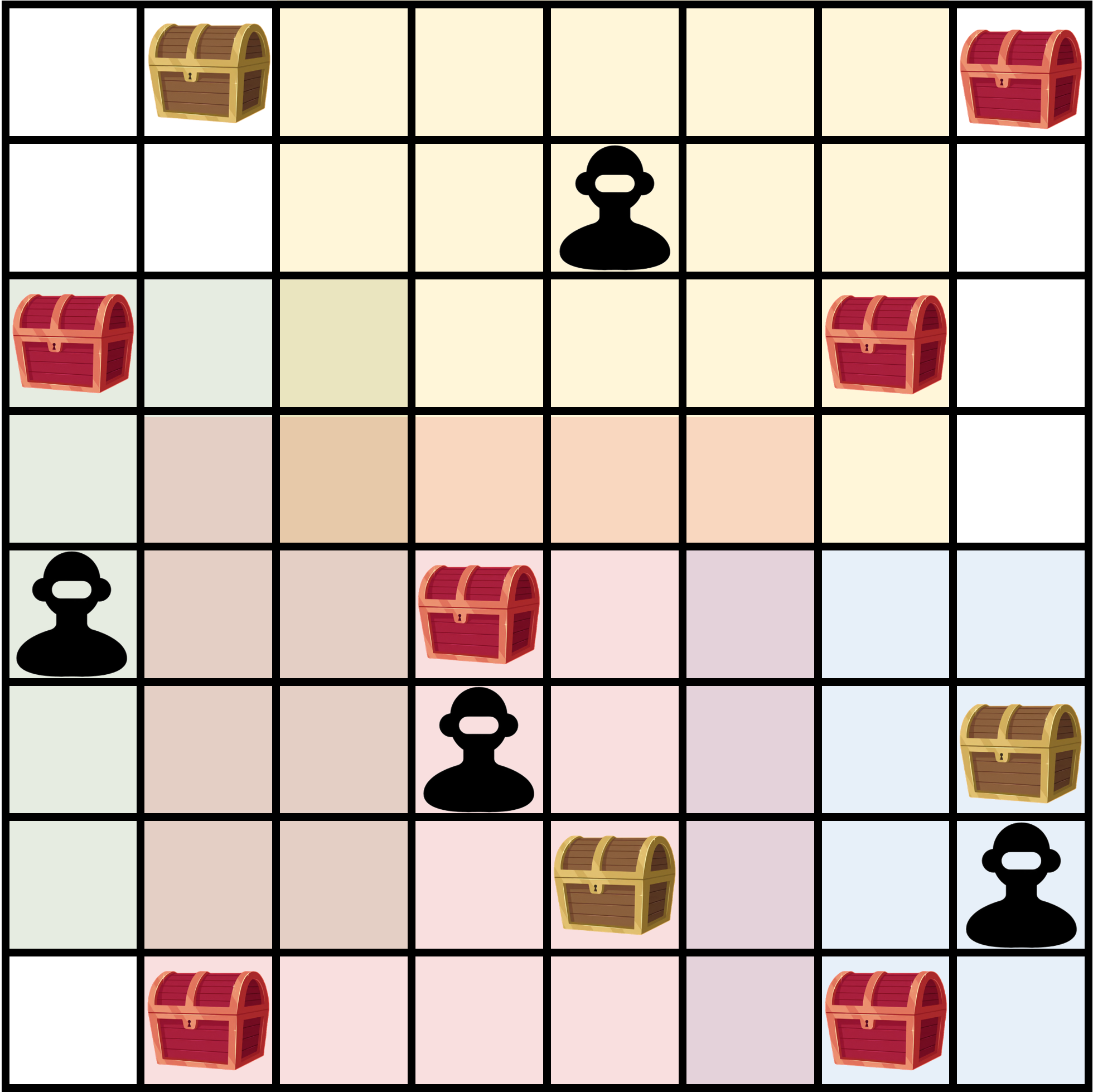}
	\caption{FT environment.}
	\label{ap-fig: ft}
\end{figure}

Find the Treasure (FT) is also a grid world game modified from \textit{ma-gym}, where four agents need to collaborate to obtain the unique treasure. The treasure is hidden inside one of the six red boxes in the environment, and the other five red boxes are empty. 
The action space of each agent is [up, down, right, left, open, pick up, stay].
One box can only be opened when both agents simultaneously perform the \textit{open} action on it; 
the items inside the box can only be collected when at least one agent performs the \textit{pick up} action at the opened box.
The observation of each agent includes: (i) is own position, (ii) the relative positions of all entities within its self-centered $5\times5$ vision filed (see the shaded area in Figure~\ref{ap-fig: ft}), and (iii) the current time step.
In addition to the unique treasure, agents can also obtain a coin from each of the three yellow boxes. This task has a limit of 50 time steps per episode and its possible \textbf{team} rewards and penalties are shown in Table~\ref{tab:ft_reward}.
\begin{table}[htbp]
	\centering
	\caption{Possible rewards and penalties \textbf{(global)} in FT task.}
	\begin{tabular}{lc}
		\toprule
		\multicolumn{1}{c}{Event} & \multicolumn{1}{c}{Reward/Penalty} \\
		\midrule
		Opening a yellow box & -1 \\
		Opening a red box & -2 \\
		Collecting a coin & 2 \\
		Collecting the treasure & 15 \\
		Step cost & -0.04 \\
		\bottomrule
	\end{tabular}
	\label{tab:ft_reward}
\end{table}

\subsection{Cleanup}
We use an open-source implementation of this environment called \textit{Sequential social dilemma games} in our experiment.
In our setting, 4 agents collaborate to collect apples in an $8\times8$ grid world.
To focus on the team coordination, we remove rotation actions, set the orientation of all agents to face \textit{up}, and disable their tagging beam, as LIO did. Therefore, an agent must move to the river side to clean waste. The action space for each agent is [up, down, left, right, clean, stay]. Each agent receives a normalized self-centered RGB image observation, with a size of $5\times5$.
To ensure that CONS and AdHocTD can work properly, we manually adjust the RGB values representing the observation center (i.e., the color of each agent) to be the same. The cleaning beam is small, 3 in length and 1 in width, so the cleaners cannot just clean the river in fixed positions.
Waste is generated uniformly in the river with a probability of $0.5$ per time step until $40\%$ of the river is covered.
When the amount of waste in the river is $0$, apples grow with a maximum probability of $0.3$ in each field grid, and this probability decays linearly to $0$ as the amount of waste increases.
Collecting one apple can bring a \textbf{team} reward of +4, and there is no punishment for the \textit{clean} action.

\section{Experimental Details}
\subsection{Implementation}
For IQL, GA-Comm and MAVEN, we adopt the implementations from \textit{MARL-Algorithms}; we also implement CONS and AdHocTD based on it. 
The Q-networks in MARL-Algorithms is a DRQN with a recurrent layer comprised of a GRU with a 64-dimensional hidden state, with a fully-connected layer before and after.
For I2Q and SEQL, we make slight modifications to the official code so that all baselines have the same experience replay buffer, optimizer, batch size and underlying network. The original AdHocTD algorithm uses tabular Q-learning and is based on the SARSA algorithm. For fair comparison, we implement it on DRQN, similar to the other algorithms. For GA-Comm, we use Central-V as the underlying algorithm, where there is a global critic network guiding the updates of actor networks for each agent. 
\subsection{Hyperparameters}
\begin{figure}[htbp]
	\centering
	\includegraphics[width=0.7\linewidth]{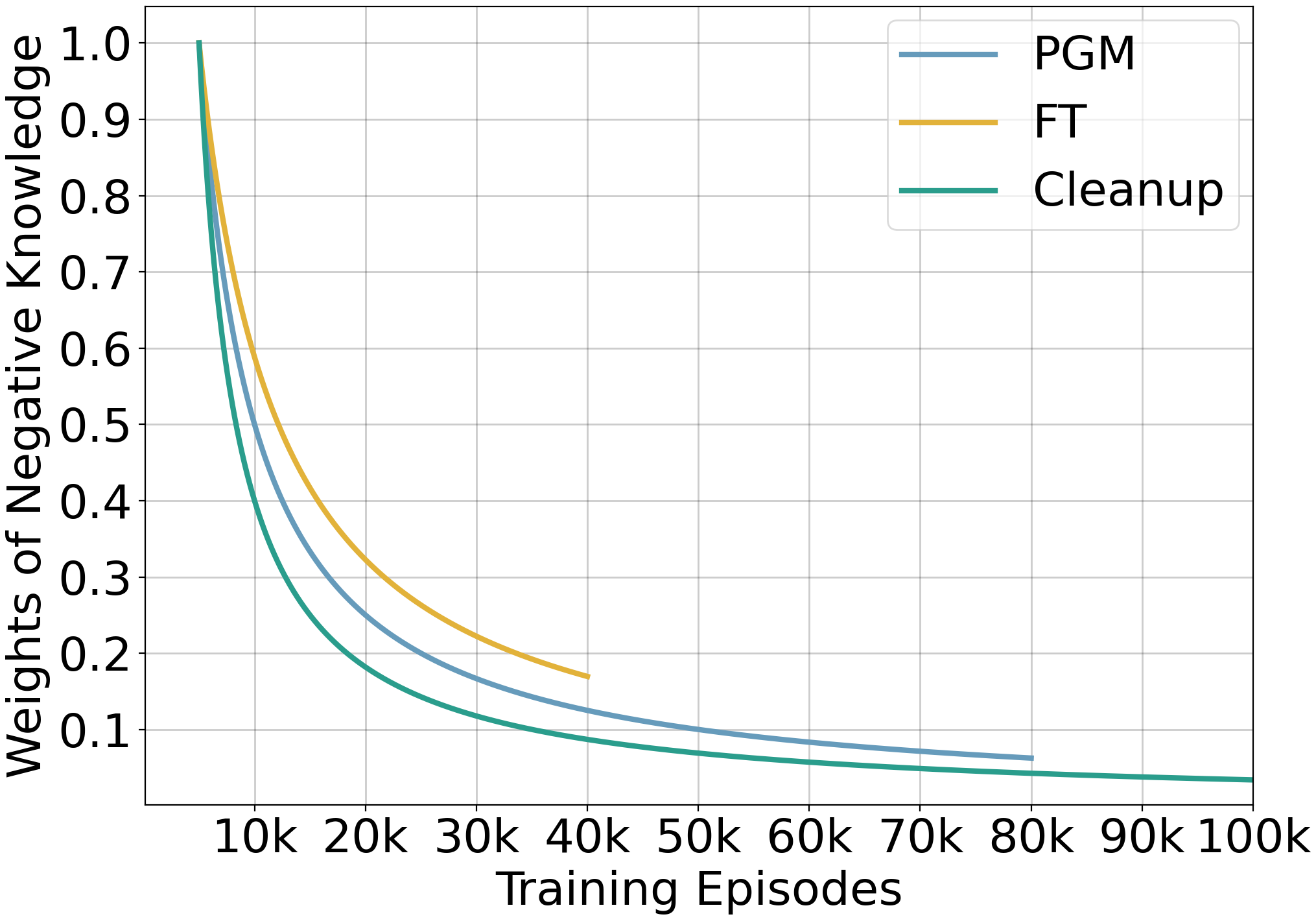}
	\caption{Changes in negative knowledge weights of CONS for PGM, FT and Cleanup.}
	\label{fig:negative knowledge}
\end{figure}

The common hyperparameters of all baselines, including CONS, are listed in Table~\ref{tab:all_parameter}.
When calculating the probability of asking for knowledge, we use the same scaling variable $\upsilon_a$, which is set to 0.5 in the PGM and FT tasks, and 0.01 in the Cleanup task. As for another scaling variable $\upsilon_g$ in AdHocTD, we adhere to the original setting and set it to 1.5 for all tasks. Moreover, we initialize all agents with budge $b_{ask}=50000$ and $b_{give}=50000(n-1)$ in CONS and AdHocTD ($n$ is the number of agents).
This is equivalent to ensuring that the agents' giving budget is always sufficient.
In addition, the unique hyperparameters specific to CONS are listed in Table~\ref{tab:cons_parameter}. The changes in negative knowledge weights of CONS in three environments are shown in Figure~\ref{fig:negative knowledge}. They are derived from Eq.8, where $e_i$ is fixed at 5000, and $\rm a$ takes values of 0, 0.3 and -0.5, respectively.

\begin{table*}[p]
	\centering
	\caption{Hyperparameters common to all algorithms.}
	\begin{tabular}{lllr}
		\toprule
		Name  & \multicolumn{2}{l}{Discription} & \multicolumn{1}{r}{Value} \\
		\midrule
		$\gamma$     & \multicolumn{2}{l}{Discounted factor} & 0.99 \\
		$|\mathcal{D}|$     & \multicolumn{2}{l}{Replay buffer size} & 10000 \\
		$n_{batch}$     & \multicolumn{2}{l}{Batch size} & 32 \\
		$lr$     & \multicolumn{2}{l}{Learning rate} & 0.0005 \\
		$t_{target}$     & \multicolumn{2}{l}{Time interval for updating the target network} & 200 \\
		$\alpha$     & \multicolumn{2}{l}{$\alpha$ value in RMSprop} & 0.99 \\
		$\epsilon$     & \multicolumn{2}{l}{$\epsilon$ value in RMSprop} & 0.00000001 \\
		$n_{hidden}$     & \multicolumn{2}{l}{Dimension of hidden layers} & 64 \\
		$\varepsilon_s$     & \multicolumn{2}{l}{Start $\varepsilon$} & 1 \\
		$\varepsilon_f$     & \multicolumn{2}{l}{Final $\varepsilon$} & 0.05 \\
		$\varepsilon$ anneal steps     & \multicolumn{2}{l}{Time-steps for $\varepsilon$ to anneal from $\varepsilon_s$ to $\varepsilon_f$} & 50000 \\
		\bottomrule
	\end{tabular}
	\label{tab:all_parameter}
\end{table*}

\begin{table*}[p]
	\centering
	\caption{The unique hyperparameters specific to CONS.}
	\begin{tabular}{lllr}
		\toprule
		Name  & Description & Source & Value \\
		\toprule
		$e_i$     & The episode of initiating knowledge sharing     & Eq.8     & 5000 \\
		\midrule
		\multirow{3}[1]{*}{$\rm a$} & \multirow{3}[1]{*}{To adjust the descent rate of the weight of negative knowledge} & \multirow{3}[1]{*}{Eq.9} & 0 (PGM)  \\
		&       &       & 0.3 (FT)  \\
		&       &       & -0.5 (Cleanup) \\
		\bottomrule
	\end{tabular}
	\label{tab:cons_parameter}
\end{table*}

\subsection{Compute Resources}

We employ different hardware configurations for running each environment, as outlined in Table~\ref{tab:hardware}.
\begin{table*}[b]
	\centering
	\caption{Hardware configurations for each environment.}
	\begin{tabular}{lll}
		\toprule
		Environment & CPU   & GPU \\
		\midrule
		Patient gold miner (PGM) & Intel(R) Core(TM) i9-10900K CPU @ 3.70GHz & NVIDIA TITAN RTX \\
		Find the treasure (FT) & Intel(R) Xeon(R) Platinum 8352V CPU @ 2.1GHz &  NVIDIA GeForce RTX 4090 \\
		Cleanup & Intel(R) Xeon(R) Gold 5118 CPU @ 2.30GHz & NVIDIA GeForce RTX 2080Ti \\
		\bottomrule
	\end{tabular}
	\label{tab:hardware}
\end{table*}

\section{Additional Results}
\begin{figure}[htbp]
	\centering
	\includegraphics[width=0.7\linewidth]{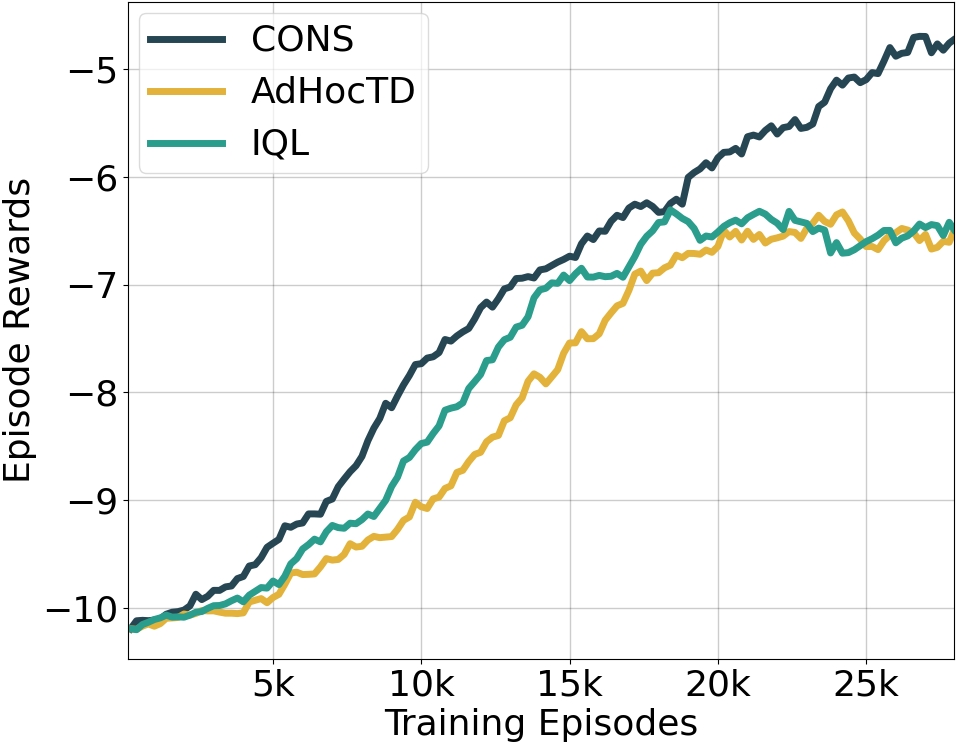}
	\caption{Additional Result for FT-20.}
	\label{ap-fig: ft20}
\end{figure}

\paragraph{Performance on Large-scale Scenario.} We believe CONS can perform well in large-scale tasks due to two reasons. First, CONS uses a decentralized training mechanism that can support parallel training of multiple agents. Second, in CONS, agents share policy-level knowledge that incurs no additional burden in training. We demonstrate the scalability of CONS on the FT task with 20 agents and gave the results in Figure~\ref{ap-fig: ft20}(a), which show the superiority of CONS over I2Q and AdHocTD in scalability.

\begin{figure}[htbp]
	\centering
	\includegraphics[width=0.7\linewidth]{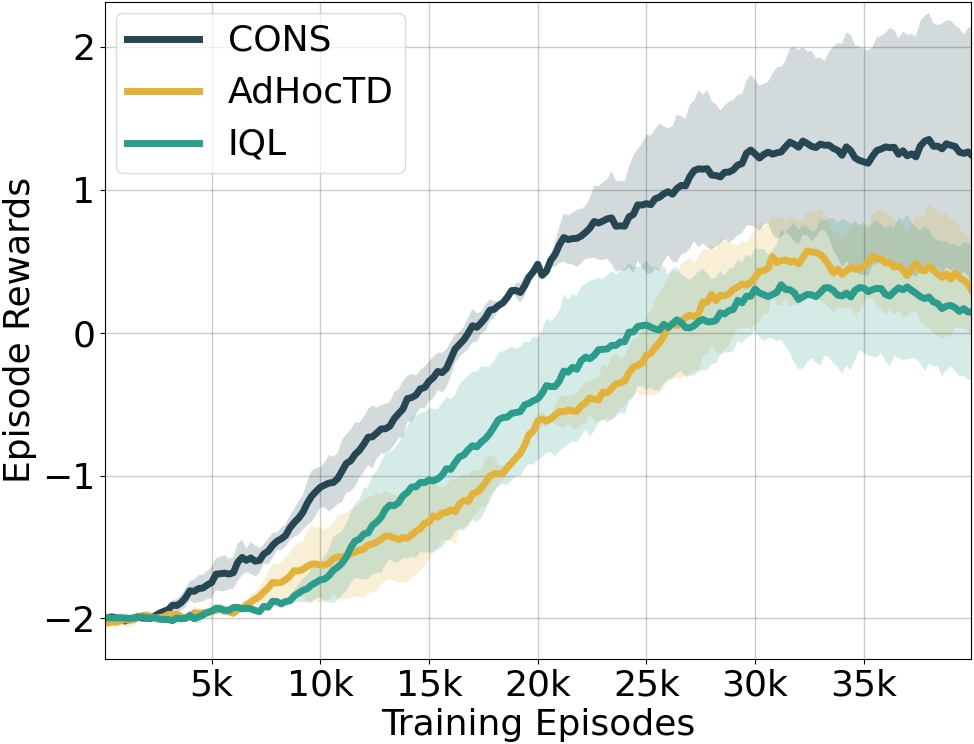}
	\caption{Additional Result for FT-multi.}
	\label{ap-fig: ftmulti}
\end{figure}

\paragraph{Performance in Scenario with Multiple Optimal Solutions.} CONS aims to avoid falling into suboptimal policies and it has no limit on the number of optimal solutions. To evaluate its performance in scenarios with multiple optimal solutions, we experiment on the FT task by adding a set of 6 green boxes as an additional optimal solution (i.e., FT-multi task). CONS still outperforms other baselines as shown in Figure ~\ref{ap-fig: ftmulti}.

\section{Pseudocode}
Algorithm~\ref{alg:2} and algorithm~\ref{alg:3} show the pseudocode of CONS teachers and students, respectively.

\begin{algorithm*}[htbp]  
	\caption{Knowledge sharing for a potential teacher $j$.}  
	\label{alg:2}  
	\begin{algorithmic}[1]
		\Require student message $m_s^i$, sharing budget $b_{give}^j$, teacher observation dictionary $C^j$, teacher Q-network $Q^j$.
		
		\If{$b_{give}^j > 0$}
		\Comment {The sharing budget is sufficient.}
		\State $o_i \gets m_s^i.o$, $n_{o_i}^i \gets m_s^i.n$, ${\rm max}Q^i(o_i,\cdot) \gets m_s^i.q$
		\Comment {Extract student information.}
		
		\State $n_{o_i}^j \gets C^j(o_i)$ 
		\Comment{Get the number of observations for $o_i$.}
		\If{$n_{o_i}^j > n_{o_i}^i$ \textbf{or} $\max Q^j(o_i,\cdot)> \max Q^i(o_i,\cdot)$}
		\Comment{Meet the requirements to share knowledge.}
		
		\State $\pi^j \gets {\rm softmax}(Q^j(o_i,\cdot))$
		\Comment{Get the action probabilities.}
		
		\State $a_b^j \gets {\rm argmax}(\pi^j)$, $a_w^j \gets {\rm argmin}(\pi^j)$
		\Comment{Get the best and worst actions.}
		
		\State $p_b^j \gets \max(\pi^j)$, $p_w^j \gets \min(\pi^j)$
		\Comment{Get the probabilities of the best and worst actions.}
		
		\State Calculate the policy confidence $\Gamma_{o_i}^j$ under $o_i$ using Eq.4.
		
		\State $\Lambda_{o_i}^j \gets \sqrt{n_{o_i}^j}\times \Gamma_{o_i}^j$.
		\Comment{Calculate its prestige under $o_i$.}
		
		\State $m_t^{ji} \gets (a_b^j, p_b^j, a_w^j, p_w^j, \Lambda_{o_i}^j)$
		\Comment{Generate the teacher message for responding to agent $i$.}
		
		\State Response to agent $i$ with $m_t^{ji}$.
		\State $b_{give}^j \gets b_{give}^j-1$ 
		\Else
		\State Remain silent.
		\Comment {Avoid inappropriate sharing.}
		\EndIf
		\Else
		\State Remain silent.
		\Comment {No budget to share knowledge.}
		\EndIf
	\end{algorithmic}
\end{algorithm*}

\begin{algorithm*}[t]  
	\caption{Action selection for a potential student $i$.}  
	\label{alg:3}  
	\begin{algorithmic}[1]
		\Require student Q-network $Q^i$, requesting budget $b_{ask}^i$, current observation $o_i$, asking probability function $P_{ask}$, student observation dictionary $C^i$, current episode number $x$, negative knowledge weight generating function $h$.
		\State Update $C^i(o_i).$
		\State $\pi^i \gets {\rm softmax}\left(Q^i(o_i,\cdot)\right)$
		\Comment {Get its original policy.}
		
		\If{$b_{ask}^i > 0$}
		\Comment {The requesting budget is sufficient.}
		
		\With {probability $P_{ask}(o_i)$} 
		\State $m_s^i.o \gets o_i$, $m_s^i.n \gets C^i(o_i)$, $m_s^i.q \gets \max Q^i(o_i,\cdot)$
		\Comment {Generate the student message for observation $o_i$.}
		\State Broadcast requesting message $m_s^i$ and waiting for teachers' responses.
		
		\If{$\{m_t^{ji}\}\neq \emptyset$}
		\Comment {Receive at least one teacher messages.}
		
		\State $\rm {getKnowledge} \gets \rm \bf {False}$
		\Comment {Variable used to record whether the probability of at least one action is modified.}
		
		\For{$\forall a \in A $}
		\State Check all $m_t^{ji}$ to find positive and negative knowledge about $a$, i.e., $K_p^a$ and $K_n^a$. 
		\If{$K_p^a \neq \emptyset$ or $K_n^a \neq \emptyset$}
		\Comment {Agent $i$ receives knowledge about action $a$.}
		\State $w_n \gets h(x)$, $w_p \gets (1-h(x))$
		\Comment {Calculate the negative and positive knowledge weights.}
		\State Modify action $a$'s probability $p_a^i$ (i.e., $\pi^i(a)$) to $\tilde{p}^i_a$ using Eq.9.
		\If {$\tilde{p}^i_a \neq p_a^i$}
		\State $\rm {getKnowledge} \gets \rm \bf {True}$
		\EndIf
		\EndIf
		\EndFor
		\If {$\rm {getKnowledge}$}
		\State Normalizing the modified action probabilities using softmax to obtain the new policy $\tilde{\pi}^i(\cdot|o_i)$.
		\State Calculate the policy confidence $\tilde{\Gamma}_{o_i}^i$ under $o_i$ using Eq.4.
		\Comment{Preparing for Targeted Exploration.}
		\With {probability $\tilde{\Gamma}_{o_i}^i$} 
		\State $a_{share} \gets \mathop{\arg\max}\limits_{a} \tilde{\pi}^i(a|o_i)$
		\Comment{Sample the best action}
		\EndWith
		
		\With {probability $1-\tilde{\Gamma}_{o_i}^i$} 
		\State Divide $[0,1]$ into $|A|-1$ equal intervals
		\If {$\tilde{\Gamma}_{o_i}^i$ is in the $q^{\rm th}$ interval in ascending order}
		\State Remove the worst $q$ actions from $\tilde{\pi}^i$.
		\Comment{$q\in{1,2,\ldots,(|A|-1)}$}
		\State Normalize the remaining action probabilities to policy $\Pi$ to be sampled.
		\State $a_{share} \gets sample(\Pi)$
		\EndIf
		\EndWith
		\Else
		\State $a_{share}\gets {\rm None}$
		\Comment {The probability of no action being modified is equivalent to not gaining knowledge.}
		\EndIf
		\Else
		\State $a_{share} \gets {\rm None}$
		\Comment {No teacher reply.}
		\EndIf	
		\EndWith
		\With {probability $\left(1-P_{ask}(o_i)\right)$} 
		\State $a_{share} \gets {\rm None}$
		\Comment{Agent $i$ does not request knowledge.}
		\EndWith
		
		\Else
		\State $a_{share}\gets {\rm None}$
		\Comment {No budget to request knowledge.}
		\EndIf
		\If{$a_{share}\neq {\rm None}$}
		\Comment {Successfully select an action based on others' knowledge.}
		\State $b_{ask}^i \gets b_{ask}^i-1$
		\State Execute $a_{share}$.
		\Else
		\State Perform $\varepsilon$-greedy exploration.
		\Comment {Fail to select an action based on others' knowledge thus focus on exploration.}
		\EndIf
	\end{algorithmic}
\end{algorithm*}

\end{document}